\documentclass{aa}
\usepackage{graphicx}
\usepackage{latexsym}
\usepackage{amsmath}
\usepackage{amssymb}
\usepackage{lscape}
\usepackage[figuresleft]{rotating}
\usepackage{txfonts}

\newcommand{\kms}{km~s$^{-1}$}
\newcommand{\sunn}{$_{\odot}$}

\newcounter{qub}
\begin{document}

\title{\ion{H}{i} study of extremely metal-deficient dwarf galaxies.}
\subtitle{I. The Nan\c {c}ay Radio Telescope observations of twenty-two objects}

\author{%
S.A. Pustilnik\inst{1,2} \and
J.-M. Martin\inst{2,3}}

\offprints{S.~Pustilnik  \email{sap@sao.ru}}

\institute{Special Astrophysical Observatory RAS, Nizhnij Arkhyz,
Karachai-Circassia,  369167 Russia
\and
Observatoire de Paris, GEPI, 5, place J. Janssen,
    92195 Meudon Cedex, France
\and
Observatoire de Paris, USN, route de Souesmes, 18330 Nan\c {c}ay,  France
}

\date{Received 27.07.2006; Accepted  \hskip 2cm 2006}

\abstract
{}
{The goal of this study is to measure parameters of the integrated
 \ion{H}{i} emission for twenty-two dwarf galaxies with oxygen abundance
12+$\log$(O/H) in the range of 7.42 to 7.65, which are representatives of the
eXtremely Metal-Deficient (XMD) galaxy group. Some of them are expected to
be similar to the well-known candidates for local young galaxies, I~Zw~18 and
SBS 0335$-$052 that have most of their baryon mass in the form of neutral gas.
Therefore, the \ion{H}{i} 21-cm line observations are crucial to understanding
their group and individual properties.}
{The Nan\c {c}ay Radio Telescope (NRT) with the upgraded  focal
receiver was used for observations of the 21-cm \ion{H}{i}-line.
This permitted the detection of the faintest sources with rms of
$\sim$1~mJy per 10.5~\kms\  resolution element.}
{For eighteen detected galaxies we
present the parameters of their integrated \ion{H}{i} line emission and
describe the data on individual objects in more detail. For four
undetected XMD galaxies, we give  upper limits on their $M$(\ion{H}{i}).
For 70\% of the twenty studied non-LSB (low surface brightness) XMD galaxies,
we find evidence (both from \ion{H}{i} and optical data) for
their interaction with neighboring objects.
In the brief discussion of the group \ion{H}{i} properties of the observed
subsample (the total O/H range is of 0.23 dex, or a factor of 1.7), we
underline
the broad distributions of the \ion{H}{i} mass (range is of 2 orders of
magnitude), of the ratio M(HI)/L$_{\rm B}$ (of 1 order of magnitude), and
of the blue luminosity (range is of 2 orders of magnitude).
We also obtained  \ion{H}{i} parameters of six
galaxies that do not belong to the XMD sample.
As a by-product, we have detected an
\ion{H}{i}-object with V$_{\rm hel} =$188~\kms, which is probably a
part of a new high-velocity \ion{H}{i} cloud (HVC 234.3-16.8+208)
near our Galaxy, in the direction of AM~0624--261.}
{These data increase the number of XMD galaxies with known integrated
\ion{H}{i} parameters (or upper limits) by a factor of two. This
allows us to address statistical properties of this group, which will
be presented in a forthcoming paper.}

  \authorrunning{S.A. Pustilnik \& J.-M. Martin}

  \titlerunning{\ion{H}{i} in 22 extremely metal-deficient galaxies}

\maketitle

\section{Introduction}

The oxygen abundances in the great majority of actively
star-forming low-mass galaxies (BCGs - blue compact and \ion{H}{ii} galaxies)
correspond to metallicities of (0.1--0.5)~Z$_{\odot}$\footnote{Updated
Z$_{\odot}$ corresponds to 12+$\log$(O/H) = 8.66, according to the recent
data by Asplund et al. (\cite{Solar04}).}
(e.g., Terlevich et al. \cite{SCHG}; Izotov et al. \cite{Izotov92}; Ugryumov
et al. \cite{HSS-LM}; Kniazev et al. \cite{SHOC}; Salzer et al.
\cite{Salzer05}; Pustilnik et al. \cite{HSS-LM2,BCG_abun}, among others).
Some BCGs, due to their low metallicities, were originally considered as
probable young galaxy populations. However, it has been shown
that the great majority of BCGs contain a measurable number of old stars
and thus, are old systems
(e.g., Papaderos et al. \cite{Papa96}; Doublier et al. \cite{Doub99};
Cairos et al. \cite{Cairos03}).
Only a few BCGs with the lowest known metallicities (12+$\log$(O/H) of
7.12--7.54, or Z = Z$_{\odot}$/34 --  Z$_{\odot}$/13) of those with well
studied deep photometry show no evidence for stellar populations older
than one Gyr.
They include  I~Zw~18 (Papaderos et al. \cite{Papa02};
Izotov \& Thuan \cite{IZw18CMD}; \"Ostlin \& Mouhcine \cite{Ostlin05});
SBS 0335--052 E and W (Papaderos et al. \cite{Papa98}; Pustilnik et al.
\cite{SBS0335}; Izotov \& Thuan \cite{IT05W}), Tol~65 (Papaderos et al.
\cite{Papa99}),  and probably DDO~68 (Pustilnik et al. \cite{DDO68}).
These BCGs show an additional blue color excess and a luminosity excess,
which are not accounted for  by the color-metallicity
(Guseva et al. \cite{Guseva03a}) and the luminosity-metallicity relations
for other BCGs
(e.g., Pustilnik et al. \cite{Kiel03}; Kniazev et al. \cite{SDSS}).
In addition, they have very high gas mass-fractions $\mu_{\rm g}$ =
M$_{\rm g}$/(M$_{\rm g}$ + M$_{\rm star}$): up to 0.95--0.99
(van Zee et al. \cite{vanZee98}; Papaderos et al. \cite{Papa02}; Pustilnik
et al. \cite{SBS0335}). These properties can be understood if these
BCGs are young objects, with non-cosmological ages.

The number of known eXtremely Metal-Deficient (XMD, 12+$\log$(O/H)
$\leq$ 7.65) galaxies has grown to $\sim$50
during the last decade thanks to dedicated searches,
conducted mainly with  objective prism surveys of  Emission Line Galaxies
(e.g., Izotov et al. \cite{Izotov92}; Kniazev et al. \cite{Kniazev98};
Melbourne \& Salzer \cite{KISS}; Ugryumov et al. \cite{HSS-LM})
and with spectroscopy of large number of
faint galaxies from the Sloan Digital Sky Survey (SDSS,  presented, e.g., by
Kniazev et al. \cite{SDSS,SHOC} and Izotov et al. \cite{SDSS_DR3}).
Such `unevolved' objects comprise only a small fraction of known galaxies
in the nearby Universe. Most of the known XMD galaxies are of the BCG type.

Different evolutionary scenarios can lead to very low metallicity of galaxy
ISM.
They include: a) the loss of metals due to superwinds related to
powerful star formation (SF) events (e.g., Mac Low \& Ferrara \cite{MLF99});
b) primordial gas inflow or intergalactic cloud accretion (e.g., Matteucci
\& Chiosi \cite{Matteucci83}; Combes \cite{Combes05});
c) the very slow astration and related metal production, characteristic of
conditions in some LSB galaxies (Low Surface Brightness -- e.g., Legrand et
al. \cite{Legrand00}), and finally, d) truly young galaxies in which the
first SF episode took place less than $\sim$1 Gyr ago.
In particular, for a few LSB dwarfs with sufficiently bright \ion{H}{ii}
regions the measurements gave very low values of
12+$\log$(O/H) of 7.3 to 7.5 (Ronnback \& Bergvall \cite{Ronnback95};
van Zee et al. \cite{vanZee96}; van Zee \cite{vanZee00}). Several dwarf
irregular galaxies (DIG) from the Local Volume were claimed as XMD objects
long ago (e.g., Skillman et al. \cite{Skillman89}).
The recent data confirmed the XMD classification for GR~8 and Leo~A (van
Zee et al. \cite{vanZee06}), DDO~53 (Pustilnik et al. \cite{VV2}) and
Sextans A and B (Kniazev et al. \cite{Sextans}). All nearby XMD LSB dwarfs
and DIGs studied with the Color-Magnitude Diagram (CMD) method
appeared to have a substantial fraction of  stars belonging to an
old stellar population. Thus,
they are old systems. In contrast to these ``non-active'' dwarfs, the CMD
data for the second
most metal-poor BCG I~Zw~18 imply that it is probably a young galaxy
(Izotov \& Thuan \cite{IZw18CMD}; \"Ostlin \& Mouhcine \cite{Ostlin05}; this
result has been questioned, however, by  Momany et al. \cite{Momany05}).
Therefore, low ISM metallicity is a necessary but not sufficient
property to assign an object as a candidate for a young galaxy.

The high-$z$ dwarf galaxies are near still or below the limit of
detectability with the most powerful instruments (e.g., Pello et al.
\cite{Pello05}; Yan et al. \cite{Yan05}; Bouwens \& Illingworth
\cite{BI2006}). Local XMD galaxy properties are similar to those of young
dwarf galaxy  populations  at high redshift, so their detailed study
is an effective way to probe early galaxy evolution. Since the cold/warm gas
should be the main component of  baryonic  matter in unevolved galaxies,
the 21-cm \ion{H}{i} line is
one of the primary means to study their properties.

Our multi-wavelength study of the XMD galaxy sample includes long-slit
spectroscopy, optical/NIR imaging, photometry, and \ion{H}{i} 21-cm spectral
observations. We present the measurements of integrated \ion{H}{i}
parameters for 22 galaxies with 12+$\log$(O/H) = 7.42--7.65 here.
These data are used for \ion{H}{i} mapping of some XMD galaxies
at the Giant Meterwave Radio Telescope (GMRT, India).
In Sect. \ref{sample} we describe the sample. Observations and data
reduction are described in Sect. \ref{observ}. The results are
presented in Sect. \ref{results}.
In Sect. \ref{discussion} we discuss the results and draw
preliminary conclusions. The analysis of $\sim$40 XMD galaxies with
available \ion{H}{i} data is deferred to a forthcoming paper.

\section{The sample}
\label{sample}

The list of XMD galaxies was compiled by Kunth \& \"Ostlin (\cite{KO},
hereafter KO). A significant number of new XMD galaxies was added in the
recent papers cited below. Metallicities of the galaxies from the KO list,
based on old, lower accuracy determinations, were revisited, and four of them
were
excluded from the XMD group (Kniazev et al. \cite{Kniazev01}; Pustilnik et
al. \cite{Pustilnik02}).

Three of the 22 galaxies are well known. AM~0624--261 is classified as a
blue LSB galaxy (Ronnback \& Bergvall \cite{Ronnback94}, \cite{Ronnback95}).
Two others - Tol~21 and Tol~65 are BCGs from the Tololo survey (Masegosa et
al. \cite{Masegosa94}; Thuan \& Izotov \cite{TI97}; Papaderos
et al. \cite{Papa99}; Fricke et al. \cite{Fricke01}, and references therein).
For these three, only AM~0624--261 has published  \ion{H}{i} data (S/N
$\sim$4, Gallagher et al. \cite{Gallagher95}).  The remaining 19 galaxies
were found recently as XMD objects. Seven  of them are from the HSS sample
(Kniazev et al. \cite{Kniazev98}; Pustilnik et al. \cite{HSS-2,BCG_abun}),
four are from the HSS-LM sample (Ugryumov et al. \cite{HSS-LM};
Pustilnik et al. \cite{HS2134}), and one is from the SBS sample (Izotov \&
Thuan \cite{IT99}). Six more galaxies were found in the SDSS (Kniazev
et al.
\cite{SDSS}). One galaxy is from the KISS sample (Lee et al. \cite{Lee04}).

No data on \ion{H}{i} parameters of these galaxies were
known in the period of our NRT observations, except for HS 0122+0743
and SDSS J1121+0324. The former was suggested as a probable
merger, so we tried to find some additional indications in its \ion{H}{i}
profile. The latter was serendipitously detected in \ion{H}{i} at the
VLA by Hibbard \& Sansom (\cite{Hibbard03}).
The main parameters of all observed galaxies with references to the sources
are shown in Table \ref{t:Param}. Some $B$-magnitudes were derived from the
photometry presented in the Sloan Digital Sky Survey (SDSS) Data Release 4
by Adelman-McCarthy et al. (\cite{DR4}), with the use of the transformation
equation from $g$ and $r$ magnitudes as given by Smith et al.
(\cite{Smith02}).

\footnotesize{\bf {
\begin{table*}
\caption[]{Parameters of the XMD sample and additional (below the line) galaxies }
\label{t:Param}
\begin{tabular}{llccrrrcccl} \hline \hline
\\[-0.3cm]
\multicolumn{1}{c}{IAU style }& \multicolumn{1}{c}{Other~~~} & \multicolumn{1}{c}{Type~ } &
\multicolumn{2}{c}{Coord. (2000.0)} & \multicolumn{1}{c}{Angul.} &
\multicolumn{1}{c}{V$_{\rm opt}$} &  \multicolumn{1}{c}{$B_{\rm tot}$$\ddagger$} &
\multicolumn{1}{c}{$M_{\rm B}^{0}$$^*$} &  \multicolumn{1}{l}{O/H$^{\dagger}$} &
\multicolumn{1}{l}{Alternative}  \\ \cline{4-5}
~~name & name or &  & R.A. & Dec.~~~ & size$^a$~~  & \kms  & mag & mag &    & name  \\
  & prefix &  &\ \ $^h$\ \ $^m$\ \ $^s$ \ & \ $^{\circ}$~~~\ \ $'$~~~\ \ $''$~~~ & a$_{25}\times b_{25}$ &    &      &    &   \\
~~~~(1) & ~~(2) & (3)  & (4)  & (5)~~~~~ & (6)~~ & (7)~~~ & (8) & (9) & (10) & ~~~(11)             \\ \hline \\
0017+1055  &HS     & BCG  & 00 20 21.35&  +11 12 20.7 &  ...         & 5630$\pm$30   & 18.2$^{1}$ &--16.7&7.63 &                 \\ 
0122+0743  &HS     & IP   & 01 25 34.18&  +07 59 22.2 & 59$\times$30 & 2926$\pm$30   & 15.5$^{1}$ &--17.7&7.63 & UGC 993         \\ 
J0133+1342 &SDSS   & BCG  & 01 33 52.56&  +13 42 09.4 & ...          & 2599$\pm$20   & 18.1$^{5}$ &--14.4&7.60 &                 \\ 
J0205--0949&SDSS   & SmPec& 02 05 49.13& --09 49 18.1 & 96$\times$48 & 1908$\pm$18   & 15.5$^{5}$ &--16.6&7.60 & KUG~0203-100    \\ 
0624--261  &AM     & LSB  & 06 26 16.99& --26 15 56.3 & 49$\times$26 &  495$\pm$17   & 16.0$^{1}$ &--12.5&7.49 & ESO~489-056     \\ 
0846+3522  &HS     & BCG  & 08 49 40.26&  +35 11 39.2 & ...          & 2177$\pm$27   & 18.1$^{5}$ &--14.6&7.65 &                 \\ 
0937+2949  &KISSB23& dIrr & 09 40 12.85&  +29 35 28.8 & 36$\times$12 &  543$\pm$11   & 16.3$^{3}$ &--13.6&7.65 & KUG~0937+298    \\ 
0940+4025  &HS     & BCG  & 09 43 36.34&  +40 11 18.1 & ...          & 5310$\pm$40   & 18.5$^{5}$ &--16.0&7.61 &                 \\ 
1013+3809  &HS     & BCG  & 10 16 24.59&  +37 54 44.5 & ...          & 1185$\pm$20   & 16.0$^{5}$ &--15.4&7.59 & KUG~1013+381    \\ 
1033+4757  &HS     & BCG  & 10 36 25.11&  +47 41 52.3 & ...          & 1541$\pm$21   & 17.7$^{5}$ &--14.3&7.65 &                 \\ 
1059+3934  &HS     & BCG  & 11 02 09.81&  +39 18 45.3 & ...          & 2995$\pm$70   & 17.2$^{5}$ &--16.1&7.62 &                 \\ 
J1105+6022 &SDSS   & BCG  & 11 05 53.62&  +60 22 28.5 & 30$\times$18 & 1340$\pm$20   & 16.6$^{5}$ &--15.1&7.62 & SBS~1102+606    \\ 
J1121+0324 &SDSS   & BCG  & 11 21 52.80&  +03 24 21.2 & 43$\times$12 & 1223$\pm$38   & 16.9$^{5}$ &--14.6&7.62 &                 \\ 
J1201+0211 &SDSS   & BCG  & 12 01 22.32&  +02 11 08.5 & ...          &  986$\pm$20   & 17.9$^{4}$ &--13.5&7.54 &                 \\ 
1159+545   &SBS    & BCG  & 12 02 02.36&  +54 15 50.1 & ...          & 3537$\pm$40   & 18.7$^{5}$ &--14.9&7.49 &                 \\ 
J1215+5223 &SDSS   & LSB  & 12 15 46.56&  +52 23 13.9 & 72$\times$40 &  162$\pm$15   & 15.4$^{5}$ &--12.9&7.42 & CGCG 269--049   \\  
1214--277  &Tol    & BCG  & 12 17 17.09& --28 02 32.6 &  ...         & 7795$\pm$50   & 18.4$^{6}$ &--16.9&7.54 & Tol~21          \\  
1223--359  &Tol    & BCG  & 12 25 46.96& --36 14 00.6 & ...          & 2811$\pm$30   & 17.6$^{7}$ &--15.6&7.54 & Tol~65          \\  
1442+4250  &HS     & BCG  & 14 44 11.35&  +42 37 34.7 & 60$\times$12 &  660$\pm$15   & 15.6$^{8}$ &--14.9&7.63 & UGC~9497        \\  
1704+4332  &HS     & BCG  & 17 05 45.39&  +43 28 48.9 & ...          & 2076$\pm$12   & 18.4$^{8}$ &--14.5&7.55 &                 \\  
2134+0400  &HS     & BCG  & 21 36 58.95&  +04 14 04.1 & ...          & 5070$\pm$50   & 19.3$^{1}$ &--15.1&7.44 &                 \\  
2236+1344  &HS     & BCG  & 22 38 31.15&  +14 00 28.6 & ...          & 6160$\pm$20   & 17.9$^{1}$ &--16.9&7.50 &                 \\ \hline 
1059+395   &KUG    & BCG  & 11 02 00.14&  +39 19 06.4 & ...          & 3067$\pm$60   & 16.6$^{3}$ &--16.8&8.11 &                 \\  
1102+6038  &MCG    & LSB  & 11 05 34.05&  +60 22 06.6 & 57$\times$34 &   --          & 16.5:$^{5}$&--15.1&     & MCG+10-16-056   \\  
J1106+6015 &MCG    & dIrr & 11 06 47.99&  +60 15 46.8 & 51$\times$17 & 1273$\pm$~3   & 16.3$^{5}$ &--15.3&     & MCG+10-16-062   \\  
J1202+0215 &       &LSB/dI& 12 02 20.94&  +02 15 58.5 & ...          &   --          & 18.8$^{5}$ & --   &     &                 \\  
1342+4210  &HS     & BCG  & 13 44 59.49&  +41 55 05.4 & ...          & 3780$\pm$50   & 18.3$^{5}$ &--15.5&7.78 &                 \\  
2236+136   &KUG    & GPair& 22 39 21.95&  +13 52 55.8 & ...          & 5184$\pm$15   & 15.4$^{5}$ &--19.0&     &                 \\  
\\[-0.3cm]
\hline\hline
\\[-0.3cm]

\multicolumn{11}{l}{($\dagger$) -- In units 12+$\log$(O/H). Data are from Ronnback \& Bergvall (\cite{Ronnback95}); Lee et al. (\cite{Lee04}); Kniazev et al. (1998);} \\
\multicolumn{11}{l}{~~~~~~~Kniazev et al. (\cite{SDSS}); Fricke et al. (\cite{Fricke01}); Guseva et al. (\cite{Guseva03b}), Pustilnik et al. (\cite{HS2134}), Pustilnik et al. (\cite{BCG_abun});}  \\
\multicolumn{11}{l}{($\ddagger$) -- Photometric data are from: $^{1}$ Pustilnik et al. in prep.; $^{2}$ Ronnback \& Bergvall (\cite{Ronnback94}); $^{3}$ Salzer et al. (\cite{KISSB_02}); $^{4}$ Kniazev} \\
\multicolumn{11}{l}{~~~~~~~et al. (\cite{SDSS}); $^{5}$ SDSS DR4; $^{6}$ Fricke et al. (\cite{Fricke01}); $^{7}$ Papaderos et al. (\cite{Papa99});  $^{8}$ Gil de Paz et al. (\cite{GM03}) } \\
\multicolumn{11}{l}{($^a$) -- Optical diameters (in $\arcsec$, measured at the 25 $B$-mag. arcsec$^{-2}$ level);   } \\
\multicolumn{11}{l}{($^*$) -- Corrected for A$_{B}$ according to Schlegel et al. (\cite{Schlegel98}) and with the distances from Table \ref{t:HI}.  } \\
\end{tabular} 
\end{table*}
     }
 }
\normalsize

\section{Observations and reduction}
\label{observ}

The \ion{H}{i}-observations with the
Nan\c {c}ay\footnote{The Nan\c {c}ay Radioastronomy Station is part of the
Observatoire de Paris and is operated by the Minist\`ere de l'Education
Nationale and Institut des Sciences de l'Univers of the Centre National
de la Recherche Scientifique.}
radio telescope (NRT) with a collecting area of
200$\times$34.5~m are characterized by a half-power beam width (HPBW) of
3.7$^{\prime}$~(East-West) $\times$ 22$^{\prime}$~(North-South) at
declination $\delta$=0$^\circ$
(see also \verb|http://www.obs-nancay.fr/nrt|).
The data were acquired during the years 2002--2004. We used 
the new antenna/receiver system F.O.R.T. (e.g., Granet et al. \cite{Granet97};
Martin et al. \cite{FORT}) with improved overall sensitivity.
The system temperature was $\sim$35 K for both the horizontal and vertical
linear polarizations of a dual-polarization receiver.
The gain of the telescope was 1.5 K~Jy$^{-1}$ at declination 
$\delta$=0$^\circ$.
The new 8192-channel correlator was used
covering a total bandwidth of 12.5 MHz. The total velocity range covered
was about 2700~\kms, with the channel spacing of 1.3~\kms\ before smoothing.
The effective resolution after averaging of four adjacent channels and
Hanning smoothing was $\approx$10.5.~\kms.
The observations consisted of separate cycles of `ON' and `OFF' integrations,
each of 40 seconds in duration. `OFF' integrations were acquired at the
target declination, usually  with the East R.A. offset of
$\sim$15\arcmin~$\times cos(\delta)$. For a few cases, when we clearly
suspected confusion from a galaxy near an `OFF' position, we selected a
different `OFF' position to avoid possible confusion.

For the flux calibration we used a noise diode. Its power was regularly
monitored throughout the observations  by pointing at   known continuum
and line sources.
The comparisons of our measured fluxes with
independent measurements of the same objects by other telescopes
indicates the consistency of the flux scale to within 10\%.

With an rms noise of $\sim$1 to 7 mJy per resolution element
after smoothing (10.5~\kms), we obtained
a S/N ratio of 30--40 for the HI line peak flux densities $F_{\rm peak}$ 
of the brightest detected objects, while for the faintest
ones we had detections with a S/N ratio of $\sim$3--4.
Total integration times per galaxy (`ON'+`OFF') varied between 1 and 10
hours. For four of the twenty-two observed XMD galaxies, we obtained only
upper limits on their $F_{\rm peak}$, and estimated limits on their integrated
\ion{H}{i} flux.

The data was reduced using the NRT standard programs NAPS and SIR,
written by the telescope staff (see description on
\mbox{http://www.obs-nancay.fr/nrt/support}).
Horizontal and vertical polarization spectra were
calibrated and processed independently and then averaged
together. The error estimates were calculated following Schneider et al.
(\cite{Schneider86}).
The baselines were generally well-fit by a third order or lower
polynomial  and were subtracted out.
For a few nearby extended galaxies with an angular size comparable to
the NRT horizontal HPBW, a correction for resolution of the observed
\ion{H}{i} flux has been done. Having no information on the \ion{H}{i} 
spatial
distribution, we used the galaxies' optical sizes and followed the procedure
described by Thuan et al. (\cite{Thuan99}). The latter takes into account
the statistical relation between the BCGs' optical size and their
characteristic size in \ion{H}{i}.

\section{Results and preliminary analysis}
\label{results}

The \ion{H}{i}-profiles of the studied galaxies, smoothed to 10.5~\kms\
are shown in Fig.~\ref{Profiles1}.
The related parameters are presented in Table \ref{t:HI}.
The velocity profiles show significant diversity  in form and width. They are
described individually in more detail  below. In Sect. \ref{discussion}
we briefly discuss some properties of the observed galaxies as a group.

In Table \ref{t:HI} we present the following data.
In Col. (1) -- the galaxy IAU type name. In Col. (2) - the central
heliocentric velocity of the  \ion{H}{i}-profile, derived as the
mid-point of profiles at 50\% of the peak, with its rms uncertainty.
The related estimate of the distance, given in Col. (3), follows that of
Karachentsev et al. (\cite{Kara04}) for their Catalog of Neighboring
Galaxies, with the adopted value of H$_{0}$ = 72~\kms~Mpc$^{-1}$.
The observed widths of \ion{H}{i}-profiles at the levels of 50 and 20~\% of
the peak value with their r.m.s. errors are given in Cols. (4) and (5).
Column (6) gives the observed \ion{H}{i} flux (the area under the profile).
The errors for the values in Cols. 4, 5, and 6 are calculated similarly to
the method suggested by Schneider et al. (\cite{Schneider86}).
Where it is appropriate, we show in Col. (7) \ion{H}{i} flux, corrected
for the beam resolution. The respective formula (see below) was discussed in
detail, e.g., by Thuan et al. (\cite{Thuan99}). The correction for source
extension in
the north-south direction is negligible since the NRT vertical beam (FWHM =
22\arcmin) is much larger than the largest size of any of the
 target galaxies.
The correction for extension in the east-west direction  (for NRT beam
of 3\farcm7) is necessary for some of the targets with
optical diameters larger than 0\farcm5.
We follow the work by Thuan \& Martin (\cite{TM81}), and assume that neutral
gas in these galaxies is distributed like an elliptical Gaussian.  Then:

\begin{center}

F$_{c}$ = F$_{H}$ $[$1 + (a$^{2}$ sin$^{2}$ PA + b$^{2}$ cos$^{2}$ PA$)$ / $\theta^{2}$]$^{1/2}$,

\end{center}

\noindent
where $a$ and $b$ are the FWHM diameters (in arc minutes) of the assumed
elliptical distribution of \ion{H}{i}, which are adopted for BCGs/\ion{H}{ii}
galaxies (following Lee et al. \cite{Lee02HI}) as 2.0 times of
the optical sizes $a_{25}$ and $b_{25}$ in Table \ref{t:Param}.  PA is the
position angle of the major axis, and $\theta$ is the HPBW
of the Nan\c {c}ay telescope in the East-West direction.
In Col. 8 we present the logarithm of the total \ion{H}{i} mass in
solar units, derived according to the well-known formula below.

\begin{center}

$M_{HI}$ = 2.36 $\times$ 10$^{5}$ F$_{c}$ $D^{2}$,

\end{center}

\noindent
where M$_{HI}$ is in $M$\sunn, F$_{c}$ is the object's integrated flux
in Jy~\kms,  and $D$ is the distance to the object in Mpc.
Finally, in Col. 9 we show the ratio M(HI)/L$_{\rm B}$ (in solar units,
 M$_{\rm B}$\sunn\ = 5.48, or L$_{\rm B}$\sunn\ = 2.12$\times$10$^{33}$
erg~s$^{-1}$),
where M(HI) is from Col. 8, and L$_{\rm B}$ corresponds to the absolute
magnitude M$_{\rm B}$ from Table \ref{t:Param}.

Besides the target galaxies, we detected \ion{H}{i} emission from several
galaxies appearing either in the `ON' beam, or in the `OFF' beam
at the radial velocities outside the uncertainty ranges for target galaxies.
We examined the respective sky regions in NED and for many cases found a
candidate galaxy that probably appeared in our \ion{H}{i} spectrum. 
Their observed \ion{H}{i} fluxes,  corrected for the offset attenuation,
are given in Col. 7.
Below we give more details on the \ion{H}{i} emission of individual galaxies
from this program, as well as other information related to the discussion
of their properties in Sect. \ref{discussion}.

\subsection{HS 0122$+$0743 = UGC 993}

This object is considered by some authors as a galaxy pair. Its morphology
indeed suggests that we are witnessing merging of the E and W components,
which seem  already to be in contact. We aimed to find more evidence
for two different components in its \ion{H}{i} profile and probable
high-velocity tails. Previous \ion{H}{i} observations with a lower S/N
ratio (Garcia et al. \cite{Garcia94}; Lu et al. \cite{Lu93}) indicated the
separate faint component in the \ion{H}{i} profile with $\delta V$ =
\mbox{--90~\kms} relative to that of the main emission. There was a hint
in the Lu et al. (\cite{Lu93}) profile of the very low contrast broad
velocity component. In our profile, the fainter component with
$\delta V$ = --75~\kms, W$_{50}\sim$35~\kms\ is well detected,
but no broad component is visible.
The width of the \ion{H}{i} profile of this object
(W$_{50}$=50~\kms) is rather narrow for its high neutral gas mass
(M(\ion{H}{i})=2.7$\times$10$^{9}$ M\sunn). According to the optical
morphology of this system, the narrow profile is probably not due to
the object being seen close to face-on.
The recent GMRT \ion{H}{i} mapping of this system
shows clear evidence of merging in this XMD object.

\subsection{SDSS J0133$+$1342}

The galaxy is marginally detected in \ion{H}{i}. Its narrow \ion{H}{i}
profile is consistent with its low optical luminosity and a small \ion{H}{i}
mass. On the DSS-2 image this galaxy is compact (maximum extent is
$\sim$16\arcsec\  or $\sim$3 kpc), with a bright central knot. Its
morphology is very disturbed. The galaxy is almost connected by a bridge to
a $\sim$3.5-mag (in $B$-band) fainter reddish galaxy at $\sim$13\arcsec\ N,
and looks like a merging system.

\subsection{SDSS J0205$-$0949 = KUG 0203$-$100}
\label{J0205}

This object has a  double-horn  profile, typical of disk galaxies, with
W$_{\rm 20}$, corresponding to the rotation velocity amplitude of
$\sim$70~\kms,
which is consistent with its optical classification in NED as SB(s)m pec.
Optical morphology, especially in the outer parts, is rather disturbed,
suggesting significant interaction. The disk is warped on both edges.
A potential candidate for a disturber is a LSB galaxy,
3\fm5 fainter (with $M_{\rm B}^0 \sim -$12.5 if at the same distance),
at $\sim$1\farcm4 SW ($\sim$11.5~kpc in projection). This galaxy is the
only candidate within the NRT beam that might be responsible for a hint of
emission in the \ion{H}{i} profile at $\Delta V \sim$ +100~\kms.

The total $B$ magnitude of SDSS J0205$-$0949 (as recalculated from its SDSS
photometry) is 15\fm46 (not $B$=18\fm38, as in Kniazev et al. \cite{SDSS};
the latter is likely  based on the light picked up by an optical
fiber of only 3\arcsec\ aperture,  thus missing a lot of the light).
With its absolute magnitude of $M_{\rm B}^0 = -$16.64, it
is somewhat bright for a galaxy with such a low metallicity.
While it could be an analog of the luminous XMD BCG HS 0837+4717
(Pustilnik et al. \cite{HS0837}), it is worth noting that its cited
uncertainty of O/H (0.09 dex) is quite large. Hence, its XMD classification
should be confirmed with higher quality spectral data.
It is worthwhile noting the high ratio M(HI)/L$_{\rm B}$=2.57, 
which is the second  highest in this sample.

\subsection{AM~0624--261 = ESO 489-056}

The first \ion{H}{i} detection for this LSB galaxy was presented by
Gallagher et al. (\cite{Gallagher95}) with the nominal S/N ratio of 4.3.
We achieved an S/N ratio of $\sim$36.
All but one of the profile parameters
of the old and new data are consistent. Only the integrated \ion{H}{i}-flux
is higher on our data by a factor of 1.35. The most reasonable explanation
of this difference is the rather low S/N ratio of the old data.
The measured $W_{50}$ = 27~\kms\ (after smoothing to 5.2 \kms)
translates (according to Staveley-Smith et al. \cite{Staveley92}, with the
adopted isotropic velocity dispersion $\sigma =$ 10~\kms) to the disk
maximal rotation velocity of 8~\kms.
An inclination correction is applied for $i = 44^{\circ}$  (Ronnback \&
Bergvall \cite{Ronnback94}).
The total \ion{H}{i} mass of this galaxy of 1.6$\times$10$^{7}$~M$_{\odot}$
is the lowest in this subsample. However, this is compatible with what one
expects for its very low blue luminosity, which is also 
the lowest in the subsample.
The ratio $M$(\ion{H}{i})/$L_{\rm B}$ = 0.85
is typical of LSB galaxies (Schombert et al. \cite{SME01}).
We detect an additional \ion{H}{i} source, seen at V$_{\rm r}$ sufficiently
close to that of the Milky Way.
It is discussed in Sect. \ref{HVC} as a possible high-velocity cloud (HVC).

\begin{figure*}
   \centering
 \includegraphics[angle=-90,width=3.8cm, clip=]{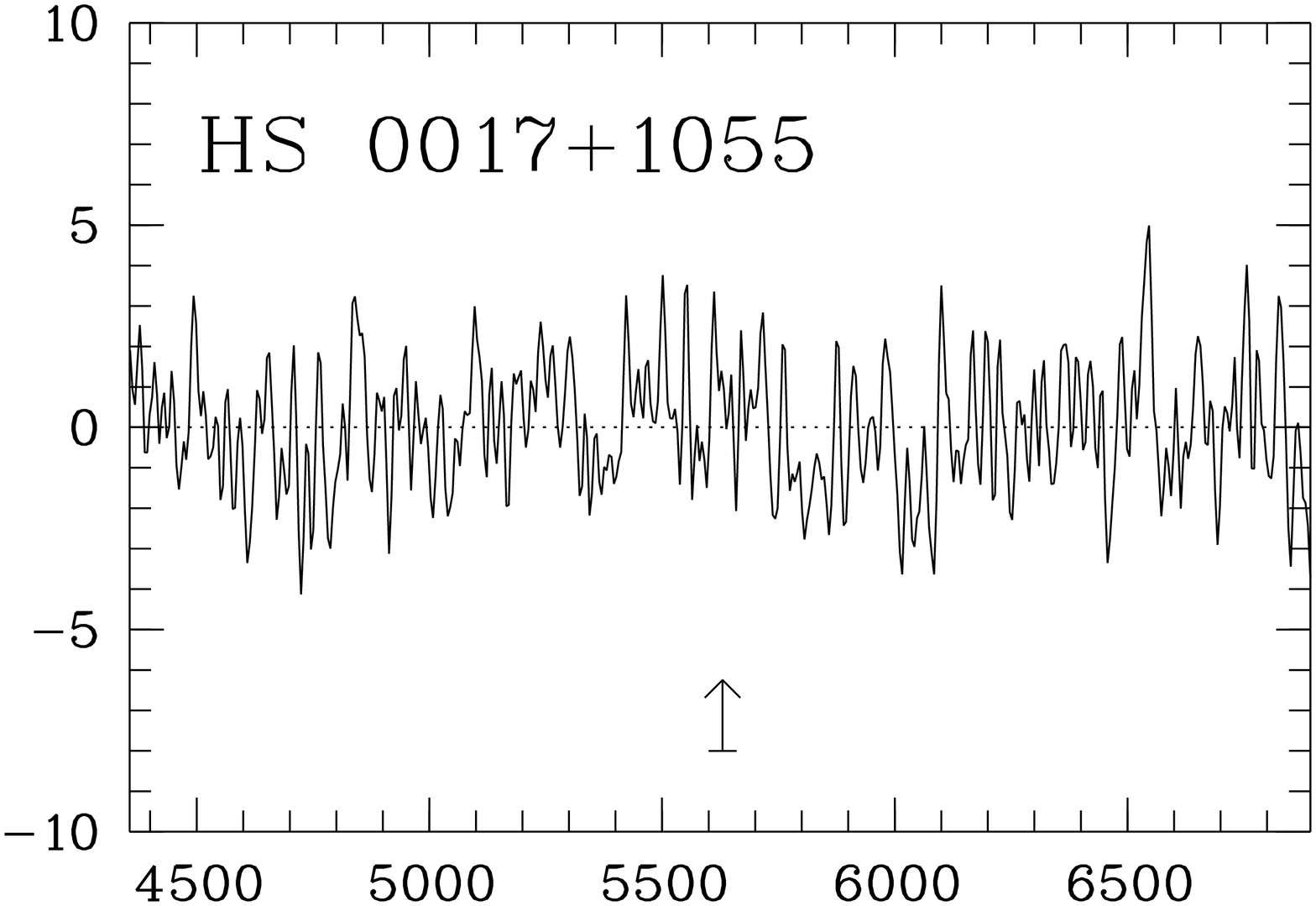} 
 \includegraphics[angle=-90,width=3.8cm, clip=]{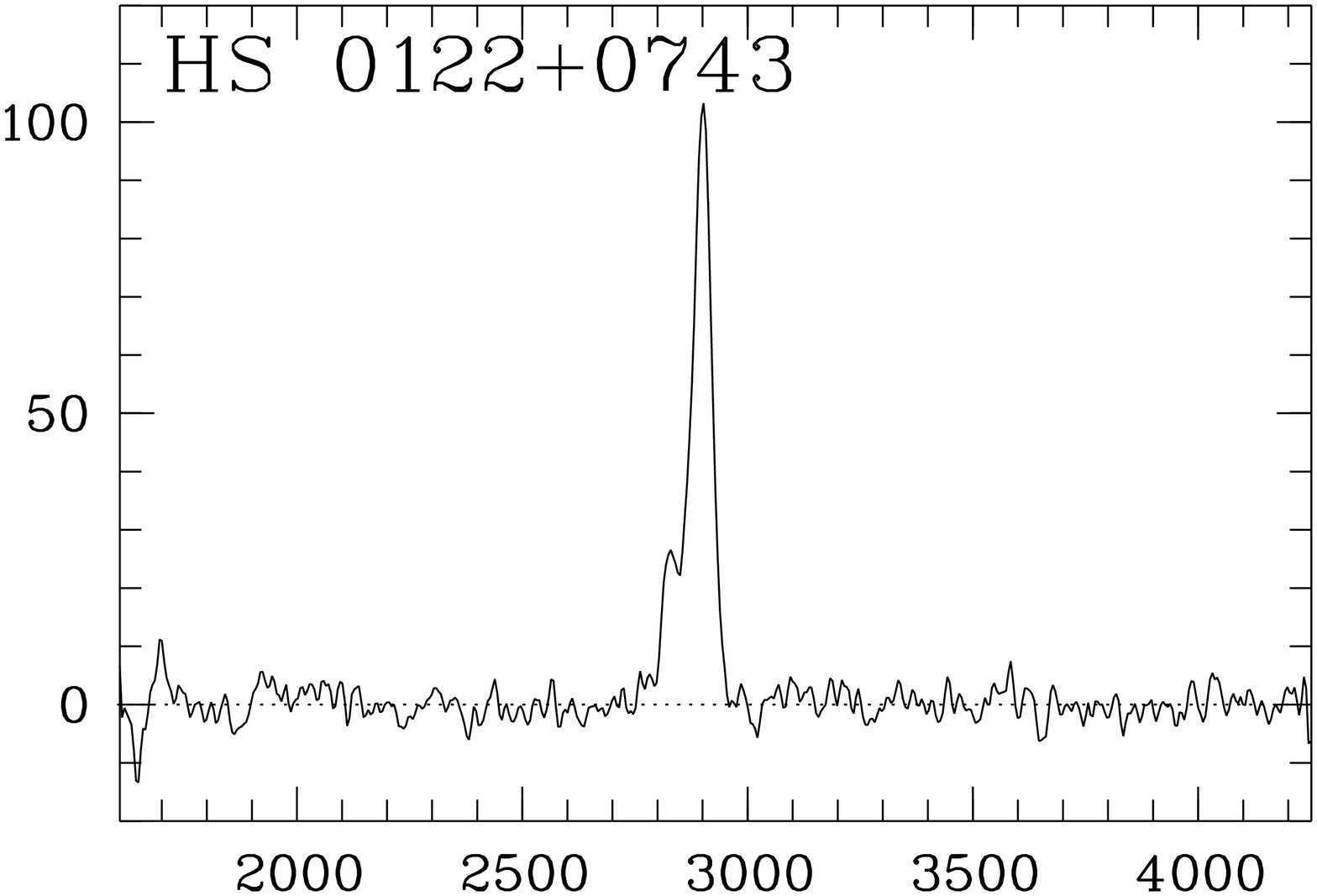} 
 \includegraphics[angle=-90,width=3.8cm, clip=]{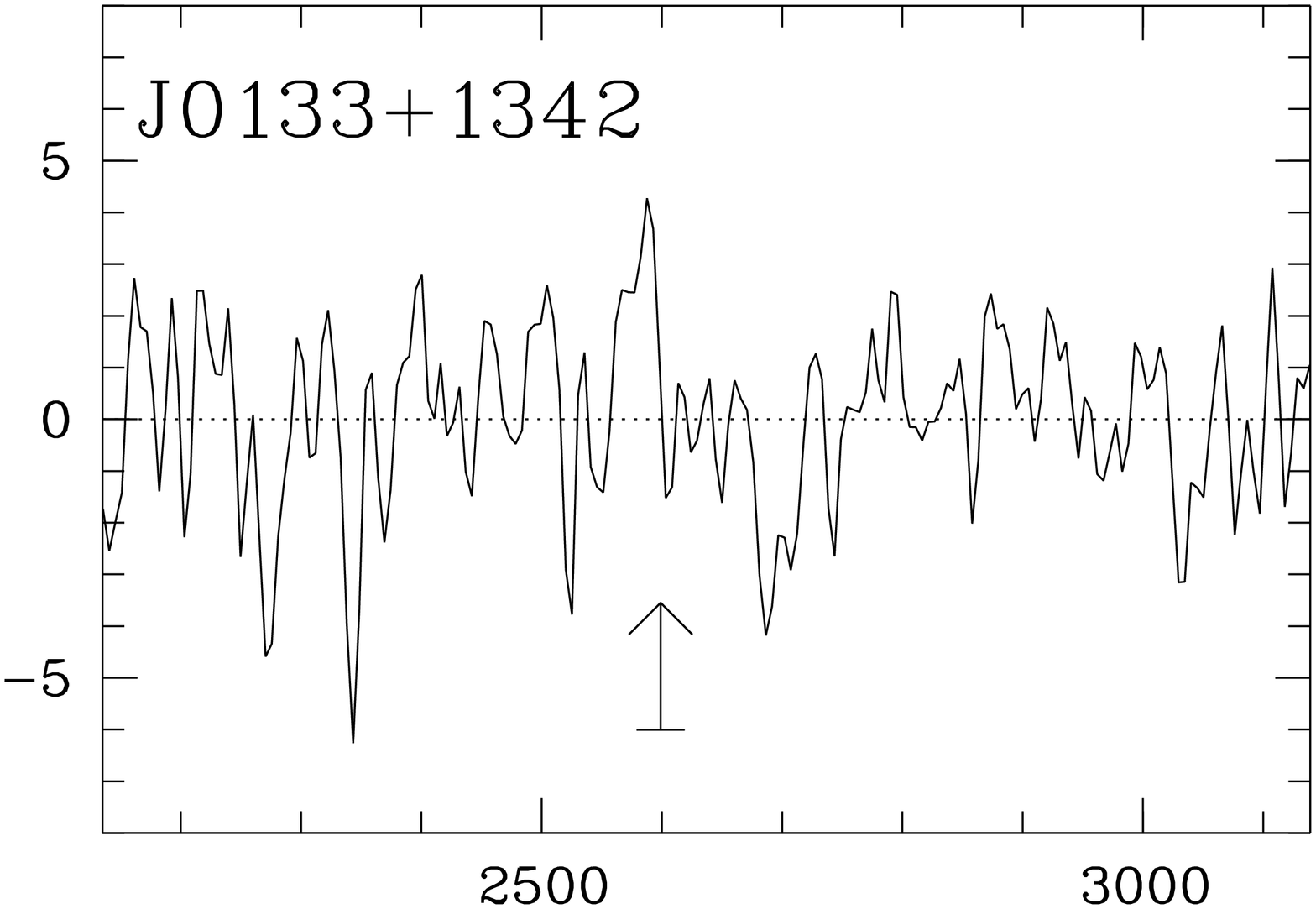} 
 \includegraphics[angle=-90,width=3.8cm, clip=]{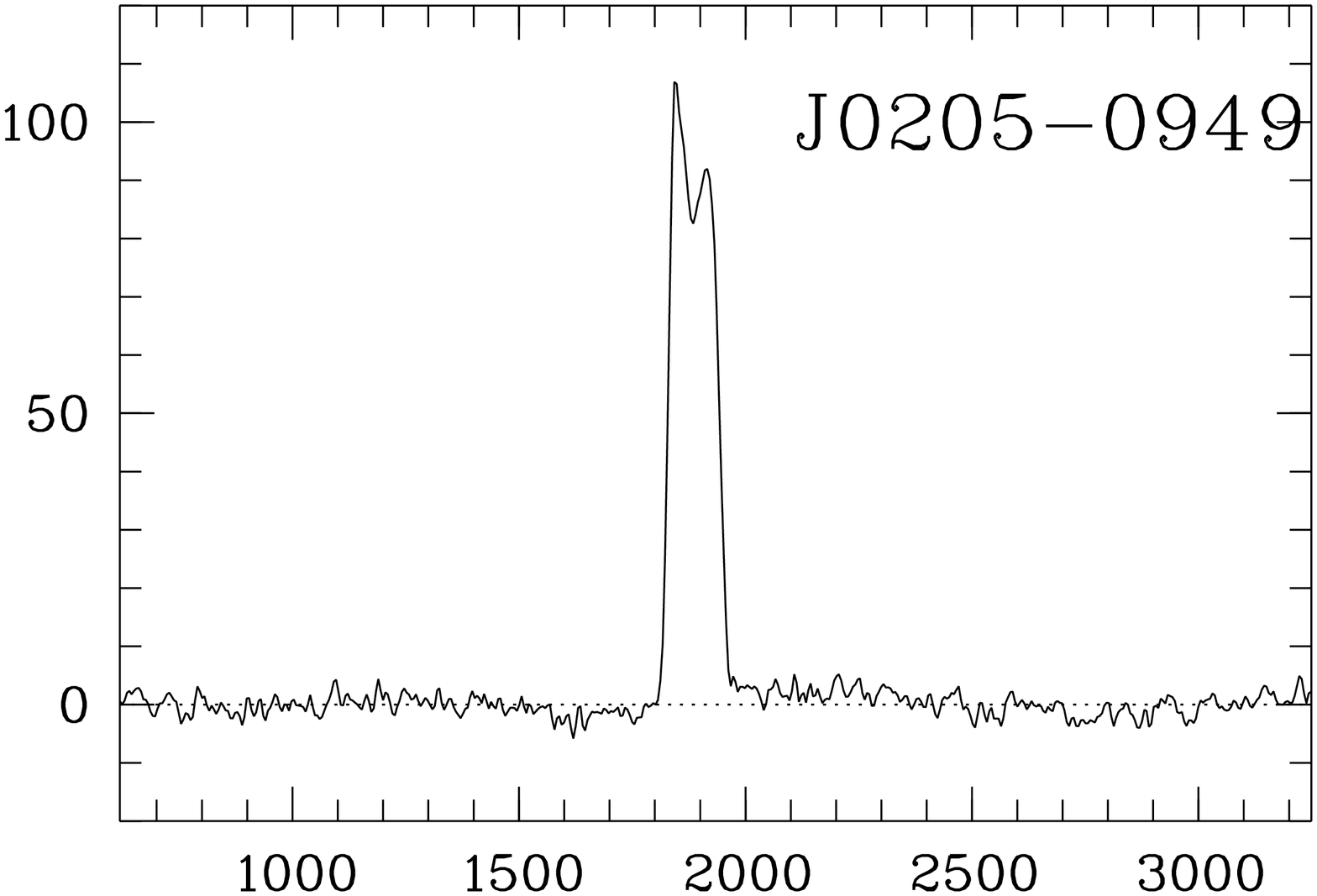} 
 \includegraphics[angle=-90,width=3.8cm, clip=]{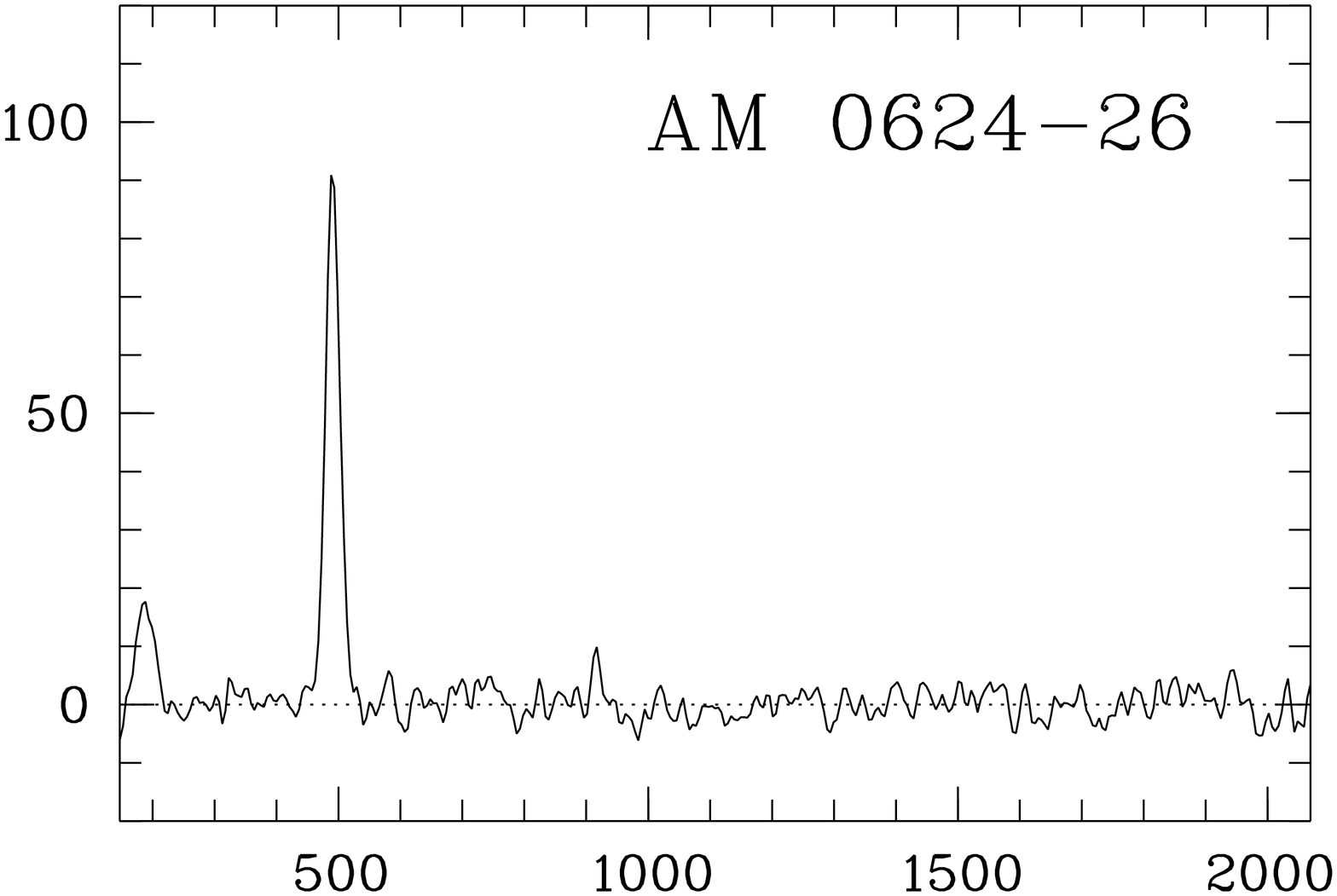} 
 \includegraphics[angle=-90,width=3.8cm, clip=]{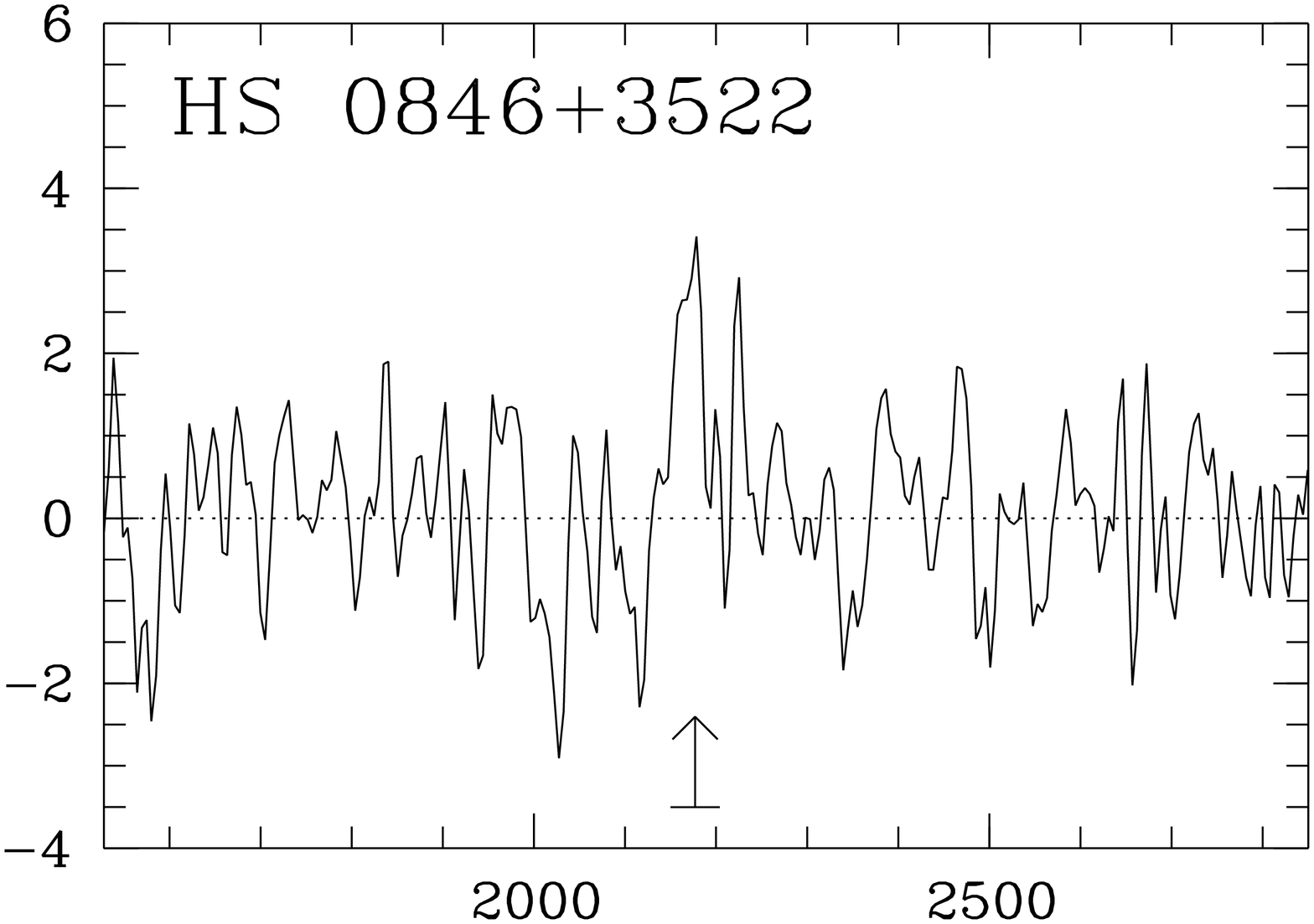} 
 \includegraphics[angle=-90,width=3.8cm, clip=]{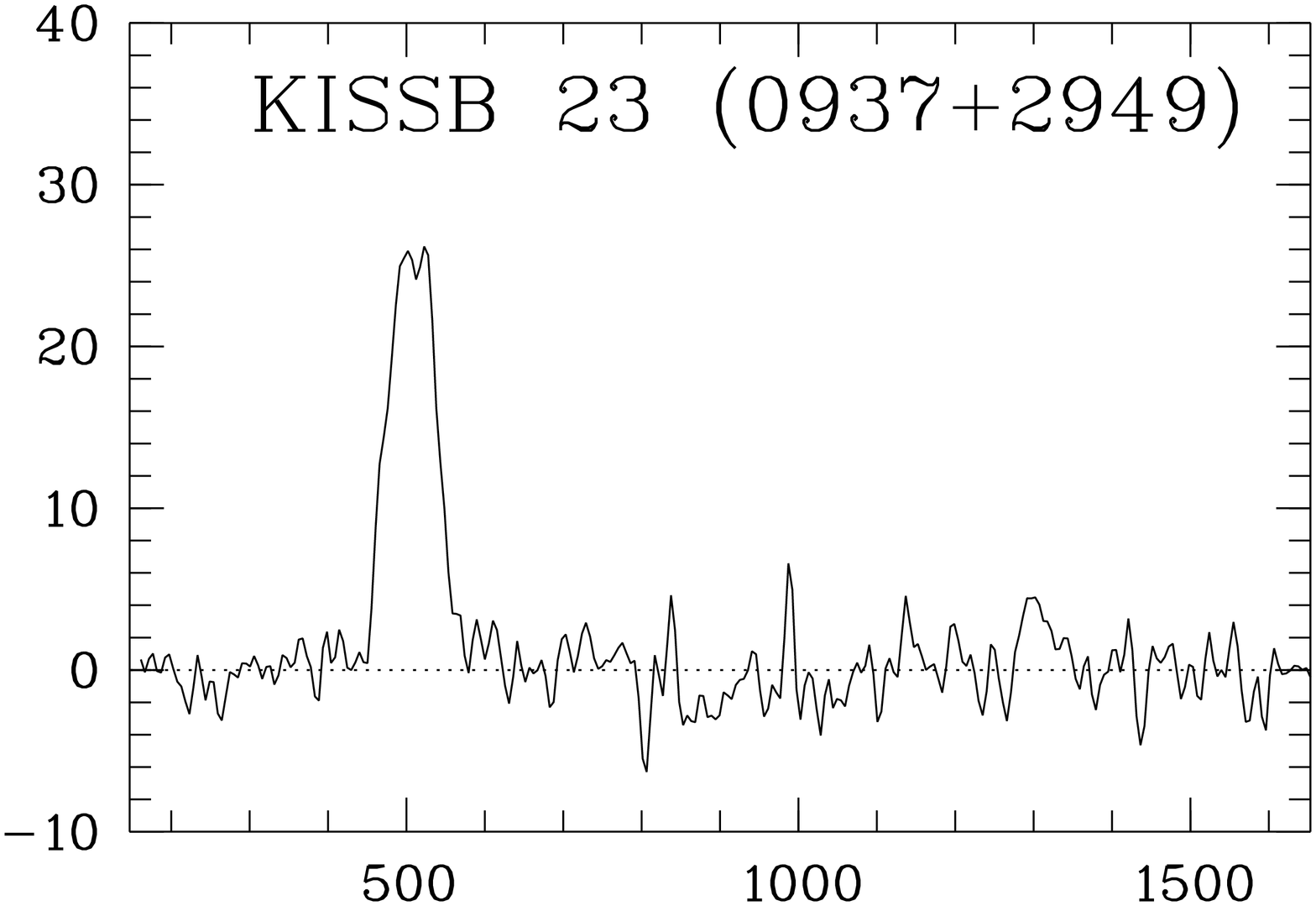} 
 \includegraphics[angle=-90,width=3.8cm, clip=]{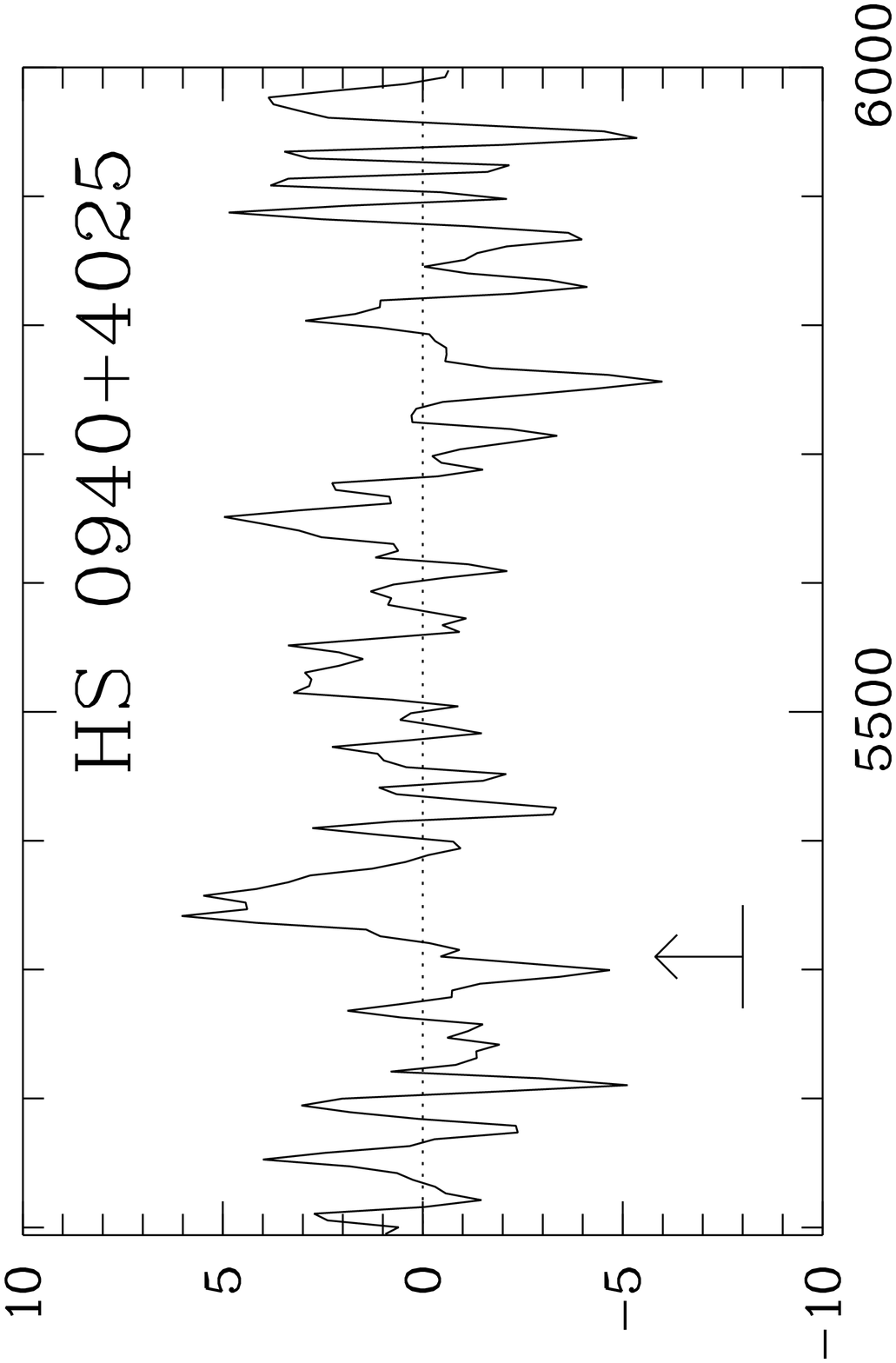} 
 \includegraphics[angle=-90,width=3.8cm, clip=]{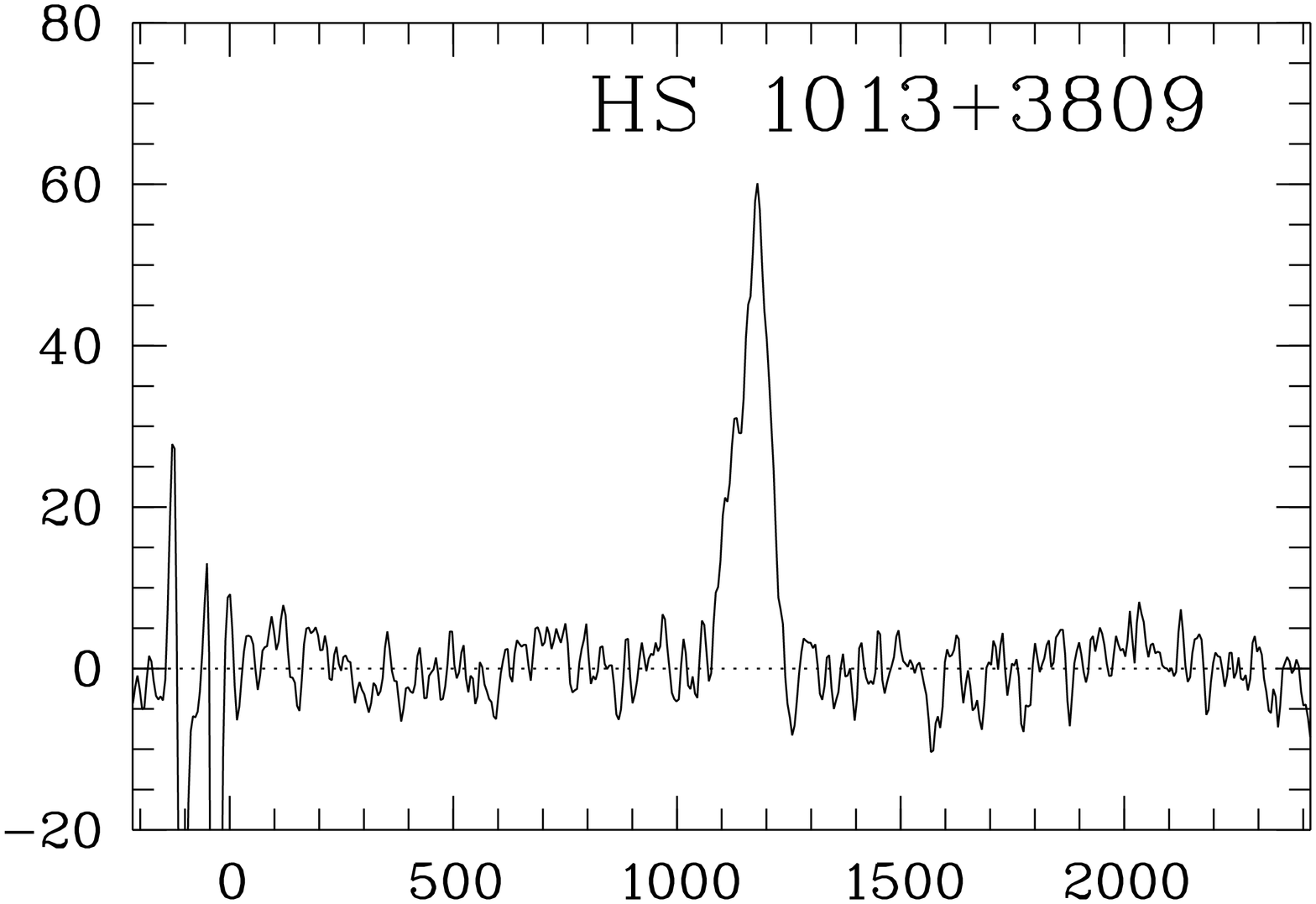} 
 \includegraphics[angle=-90,width=3.8cm, clip=]{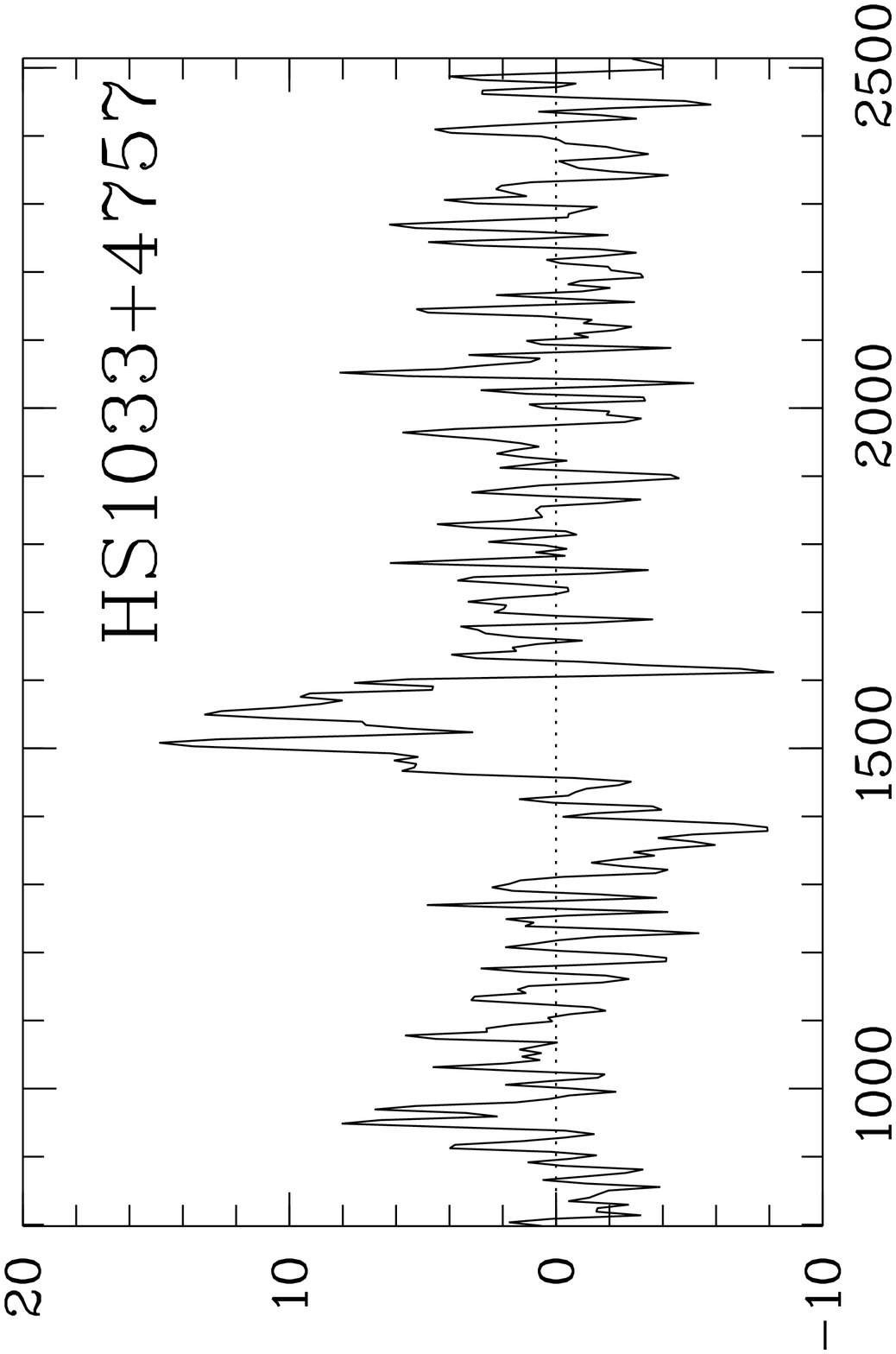} 
 \includegraphics[angle=-90,width=3.8cm, clip=]{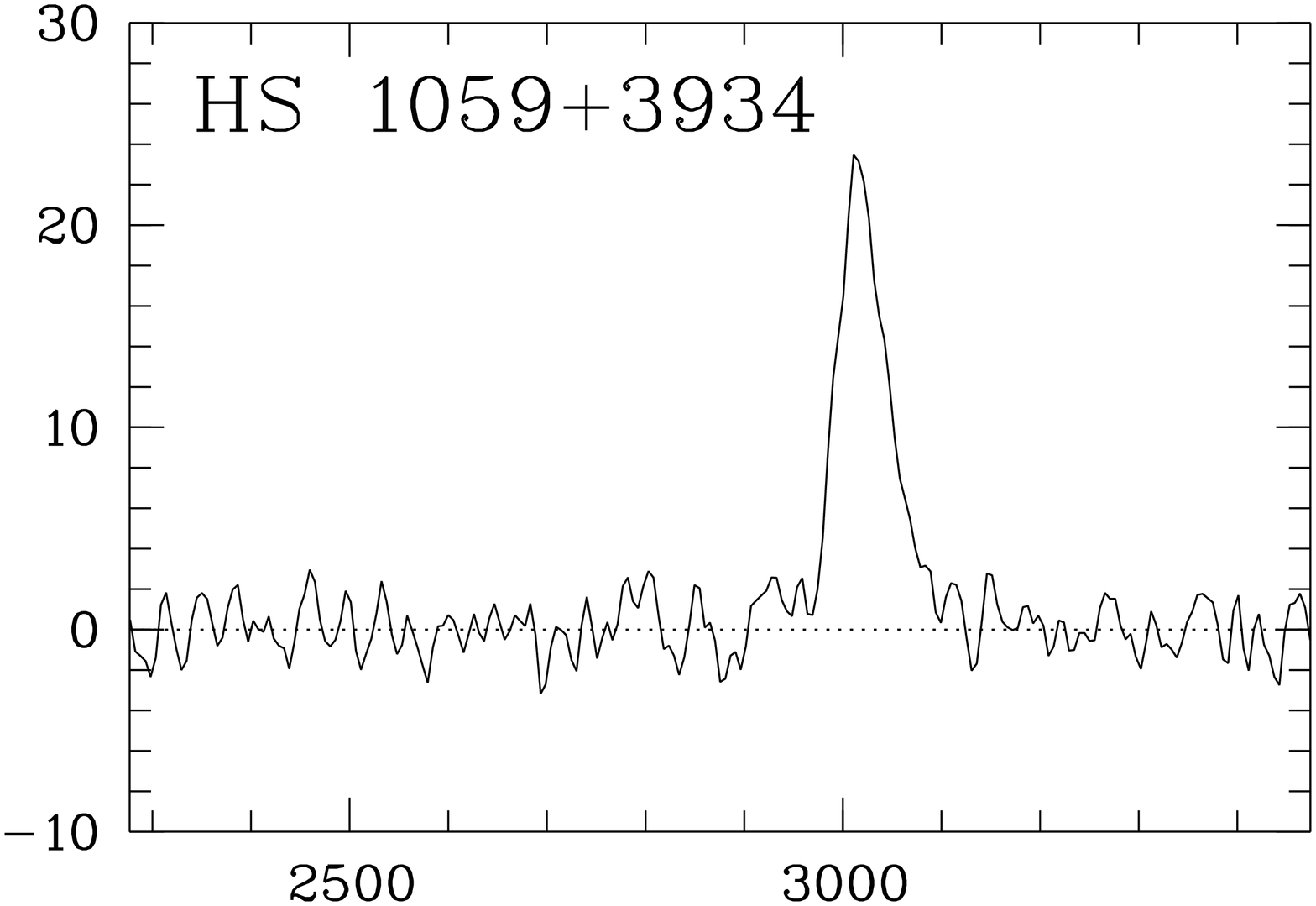} 
 \includegraphics[angle=-90,width=3.8cm, clip=]{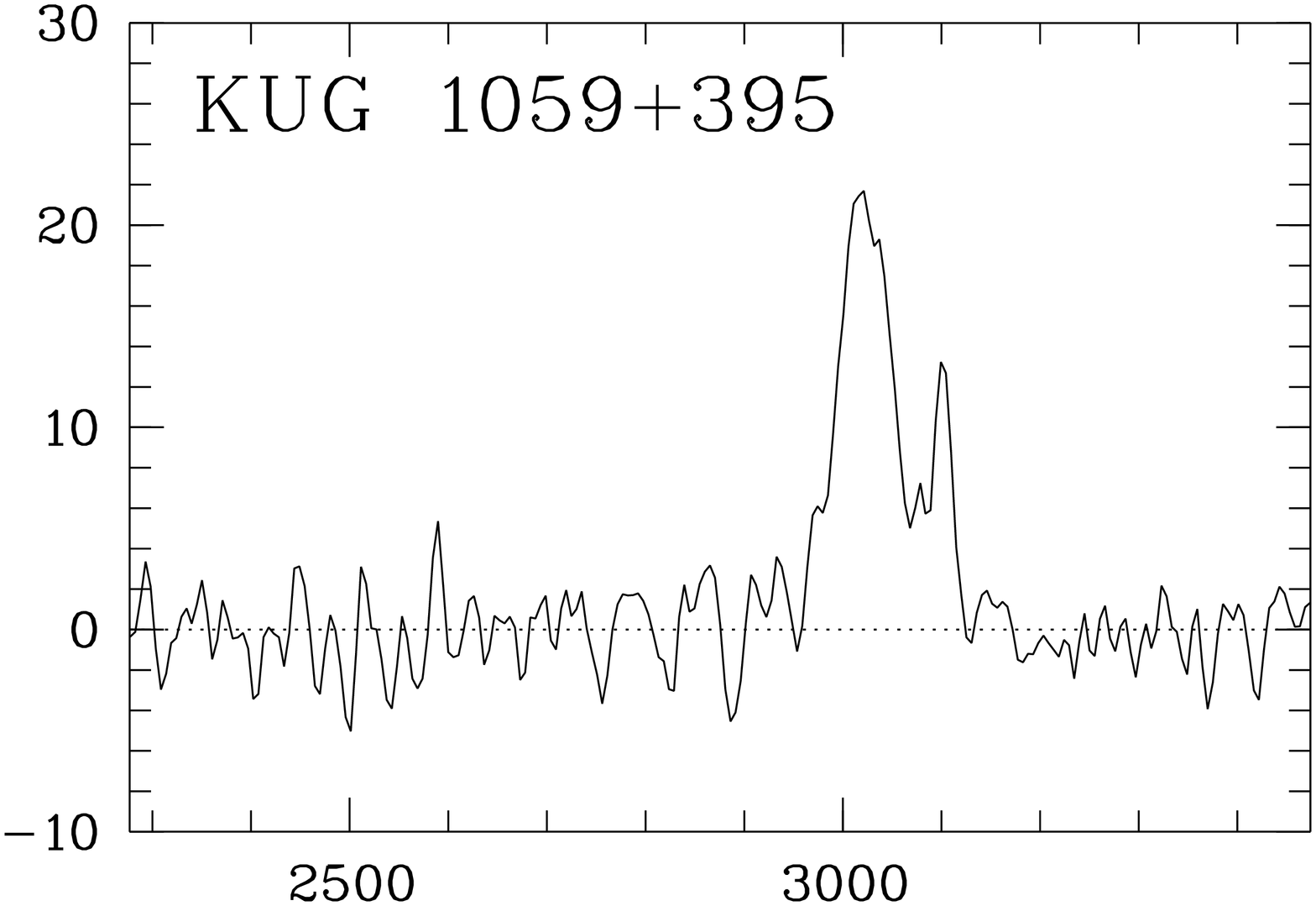} 
 \includegraphics[angle=-90,width=3.8cm, clip=]{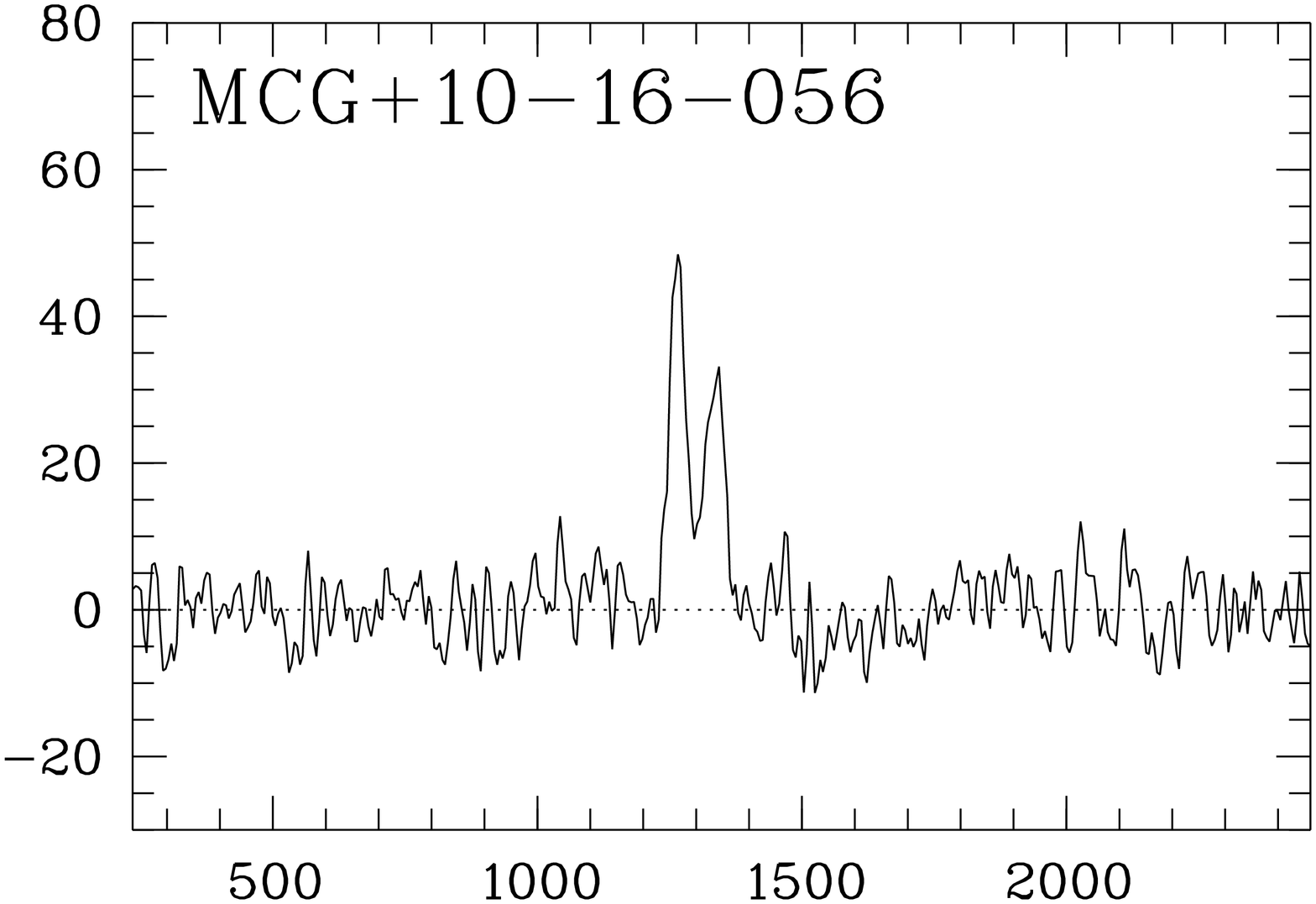} 
 \includegraphics[angle=-90,width=3.8cm, clip=]{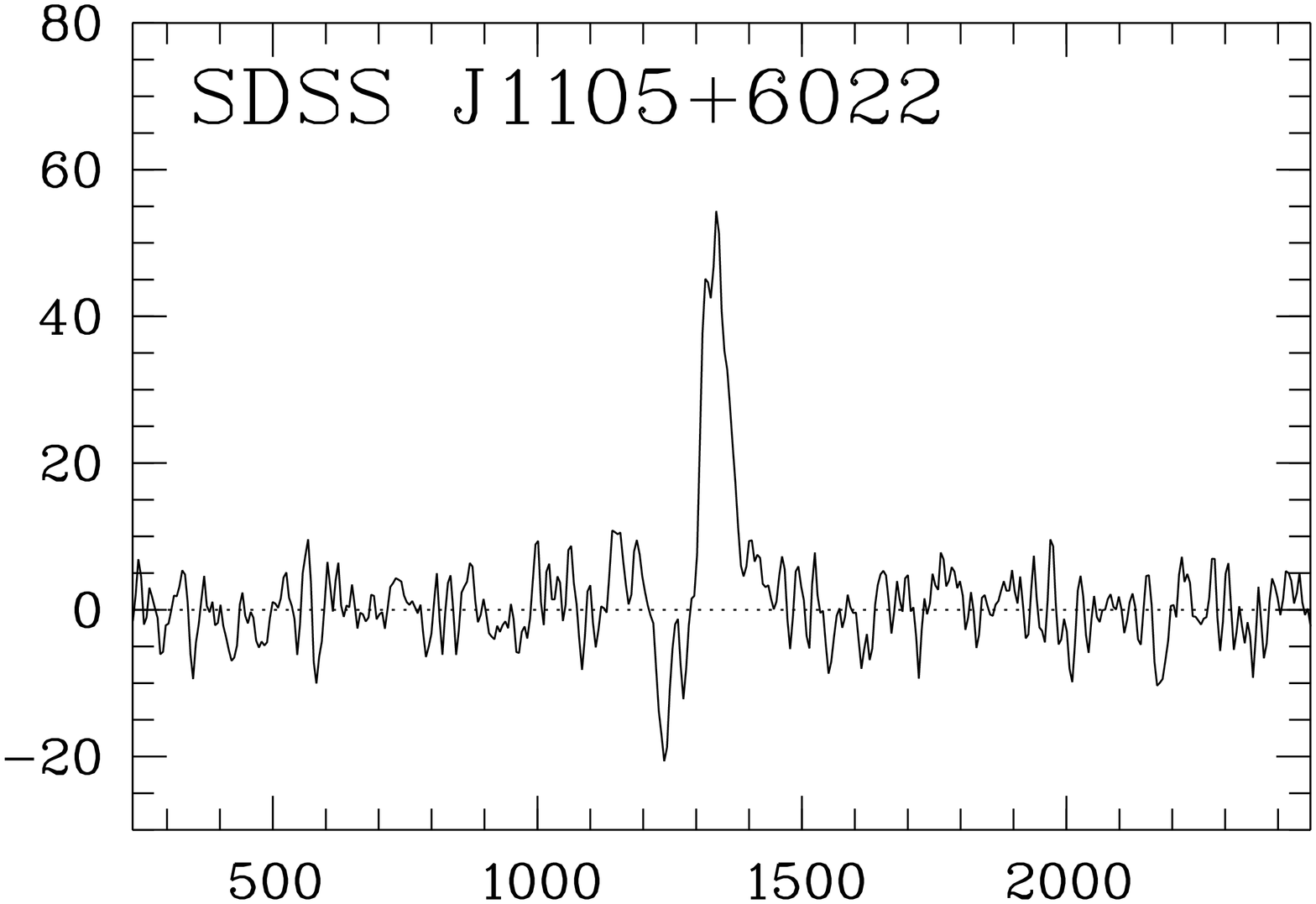} 
 \includegraphics[angle=-90,width=3.8cm, clip=]{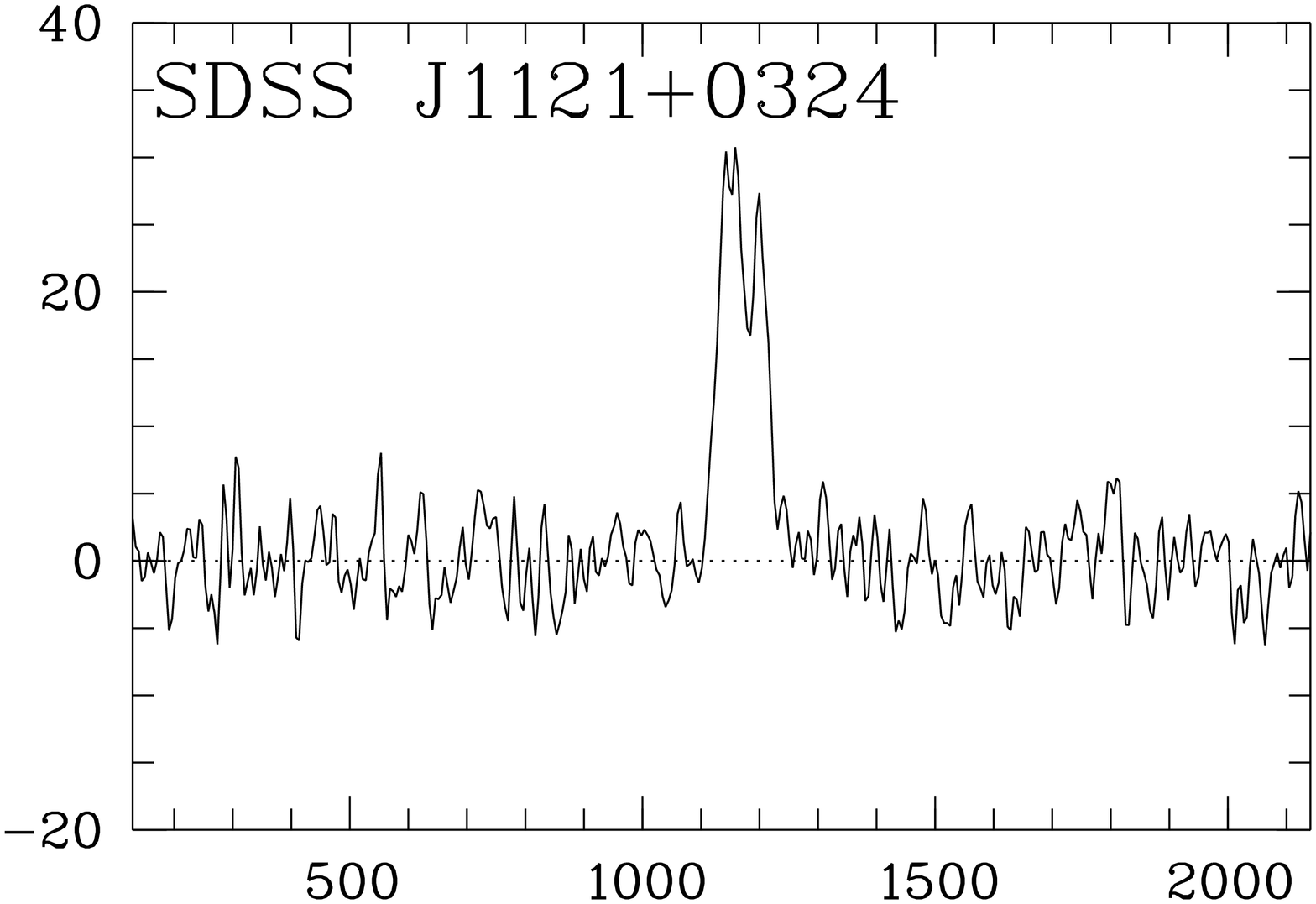} 
 \includegraphics[angle=-90,width=3.8cm, clip=]{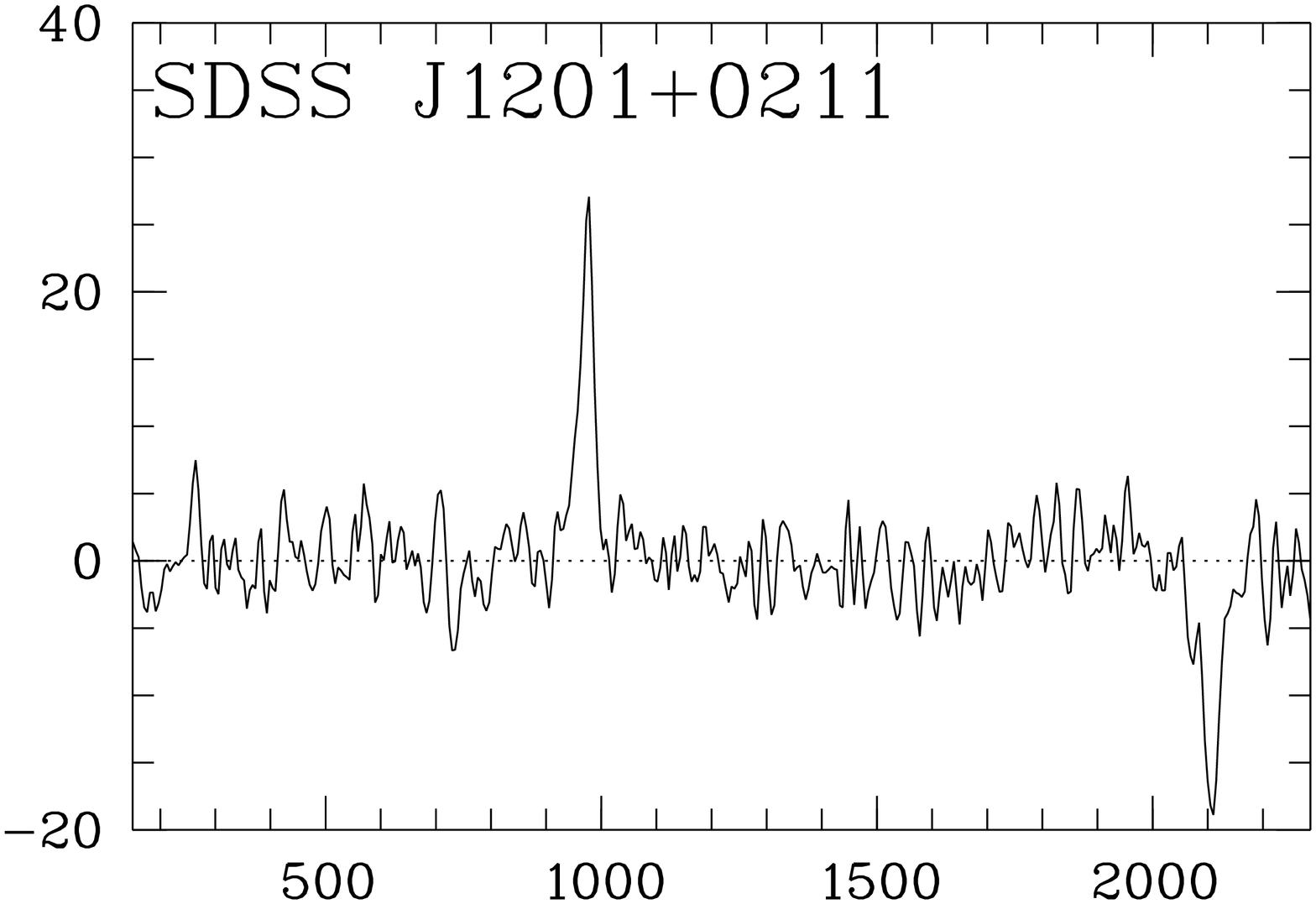} 
 \includegraphics[angle=-90,width=3.8cm, clip=]{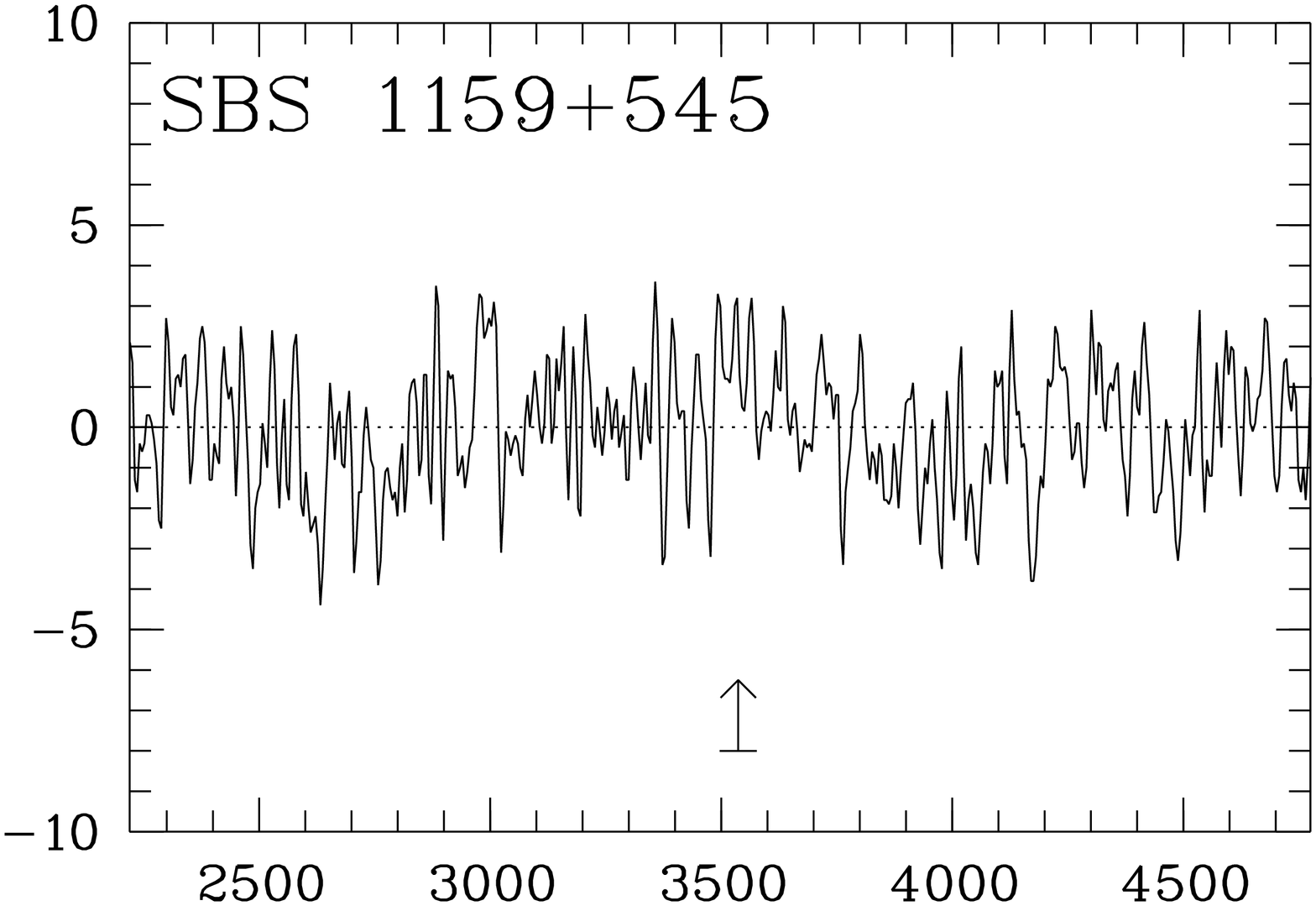} 
 \includegraphics[angle=-90,width=3.8cm, clip=]{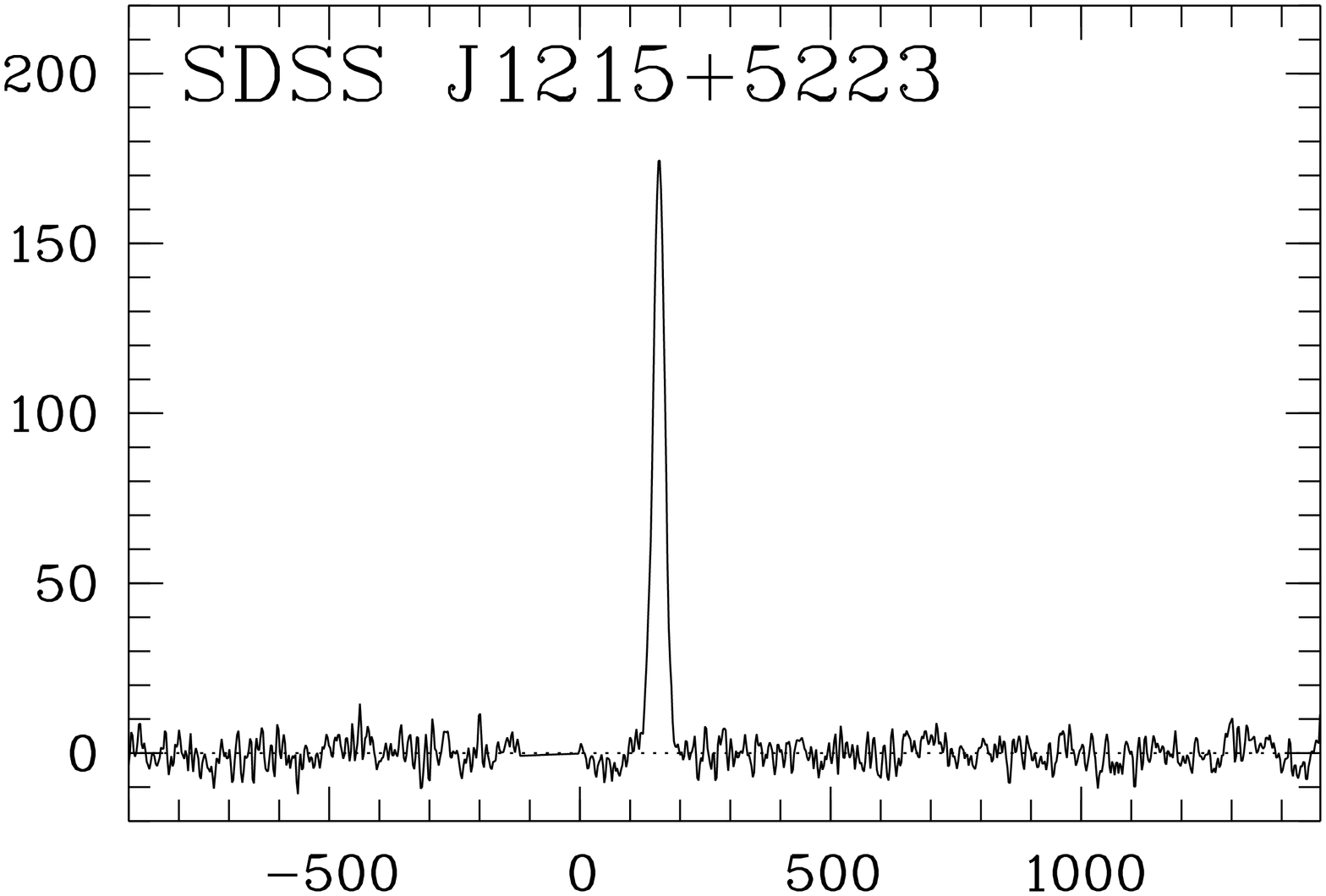} 
 \includegraphics[angle=-90,width=3.8cm, clip=]{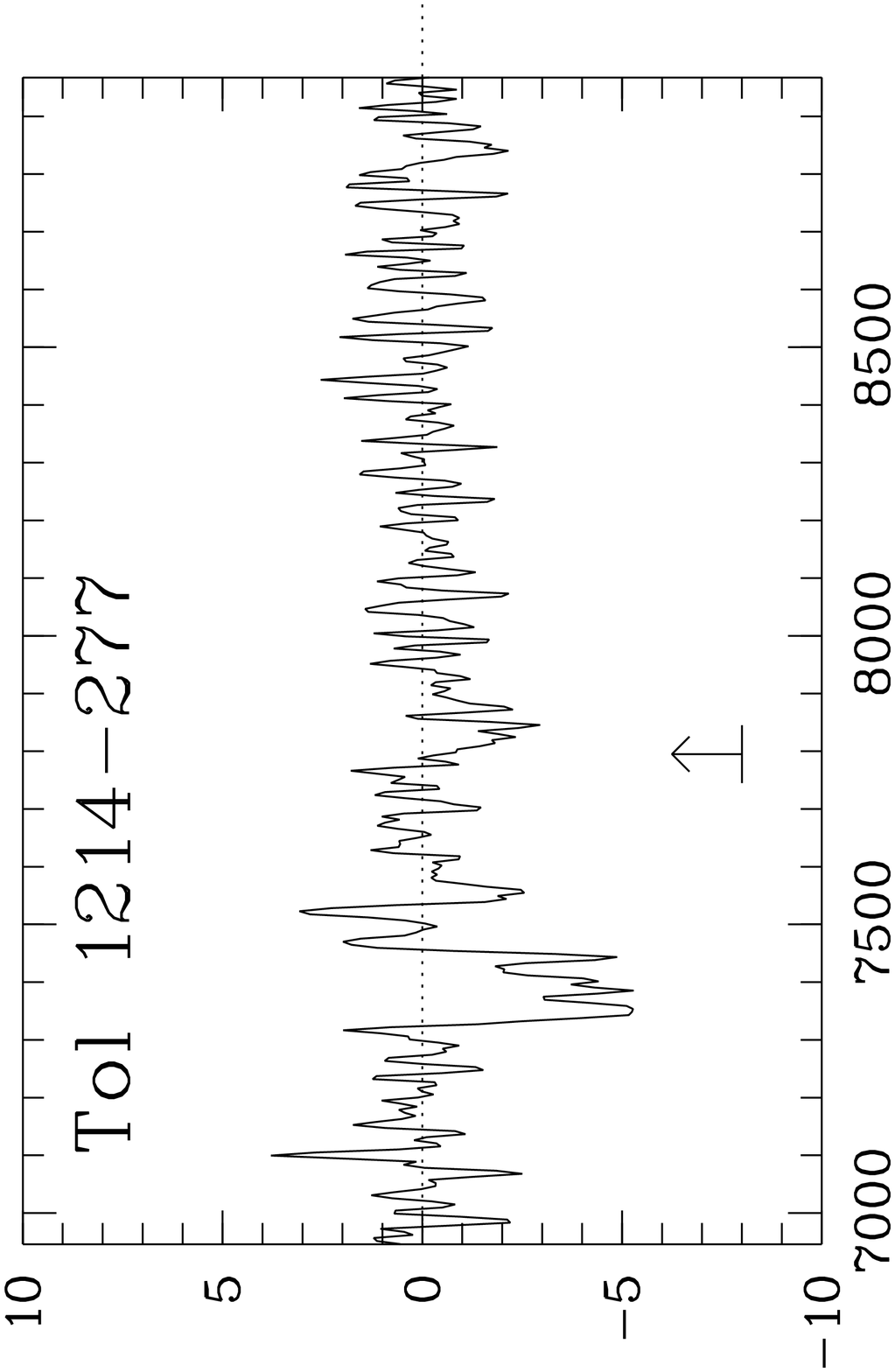} 
 \includegraphics[angle=-90,width=3.8cm, clip=]{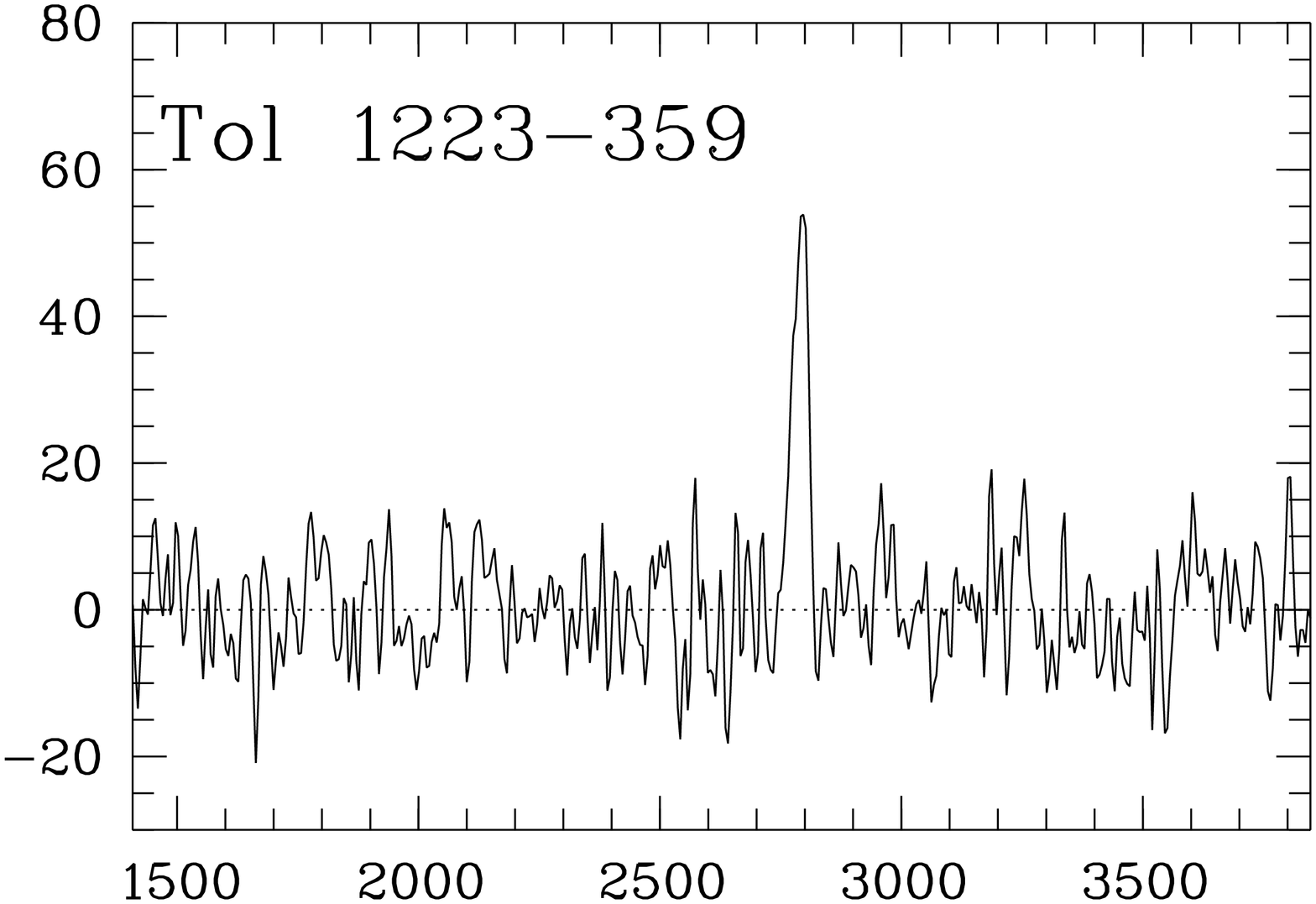} 
 \includegraphics[angle=-90,width=3.8cm, clip=]{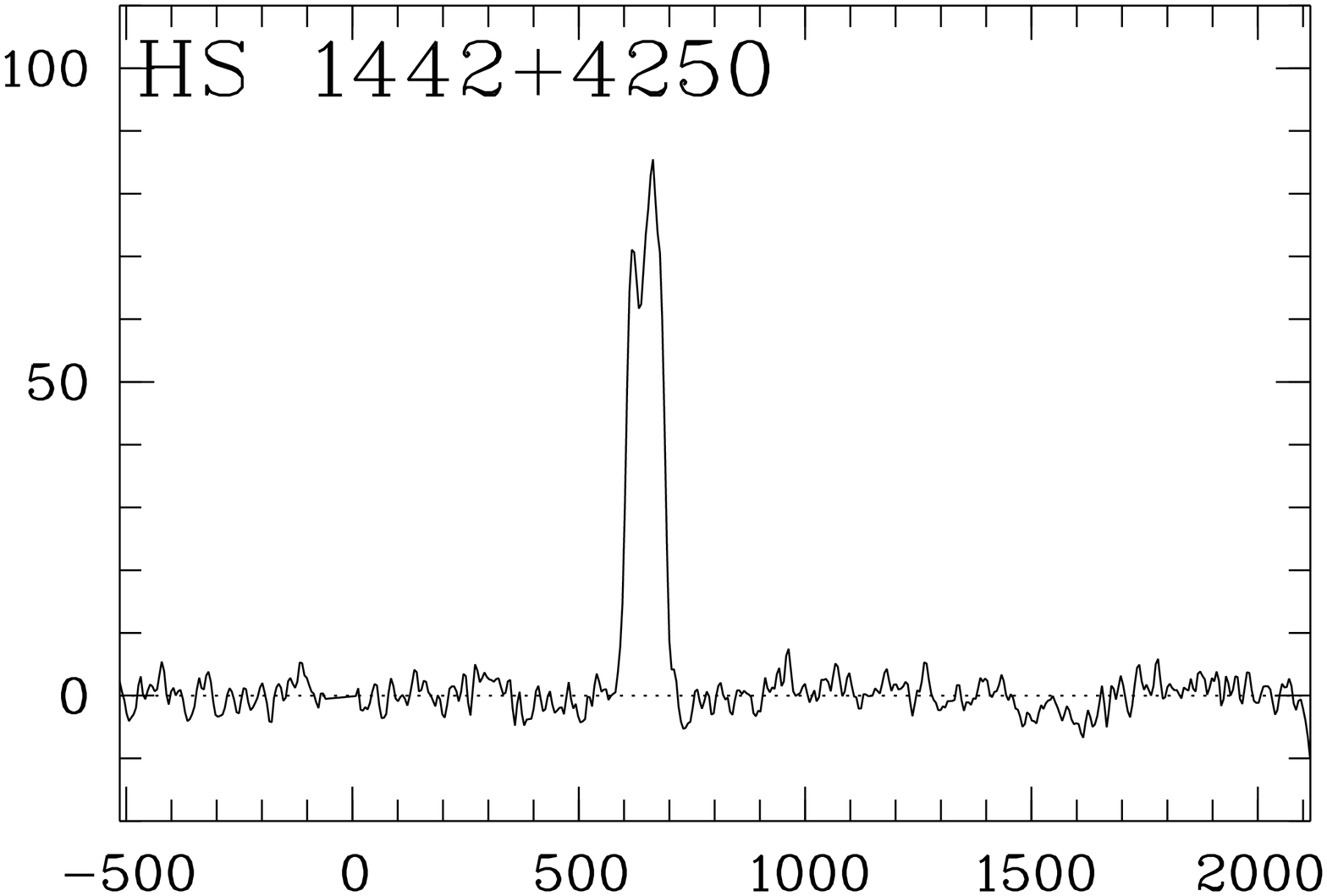} 
 \includegraphics[angle=-90,width=3.8cm, clip=]{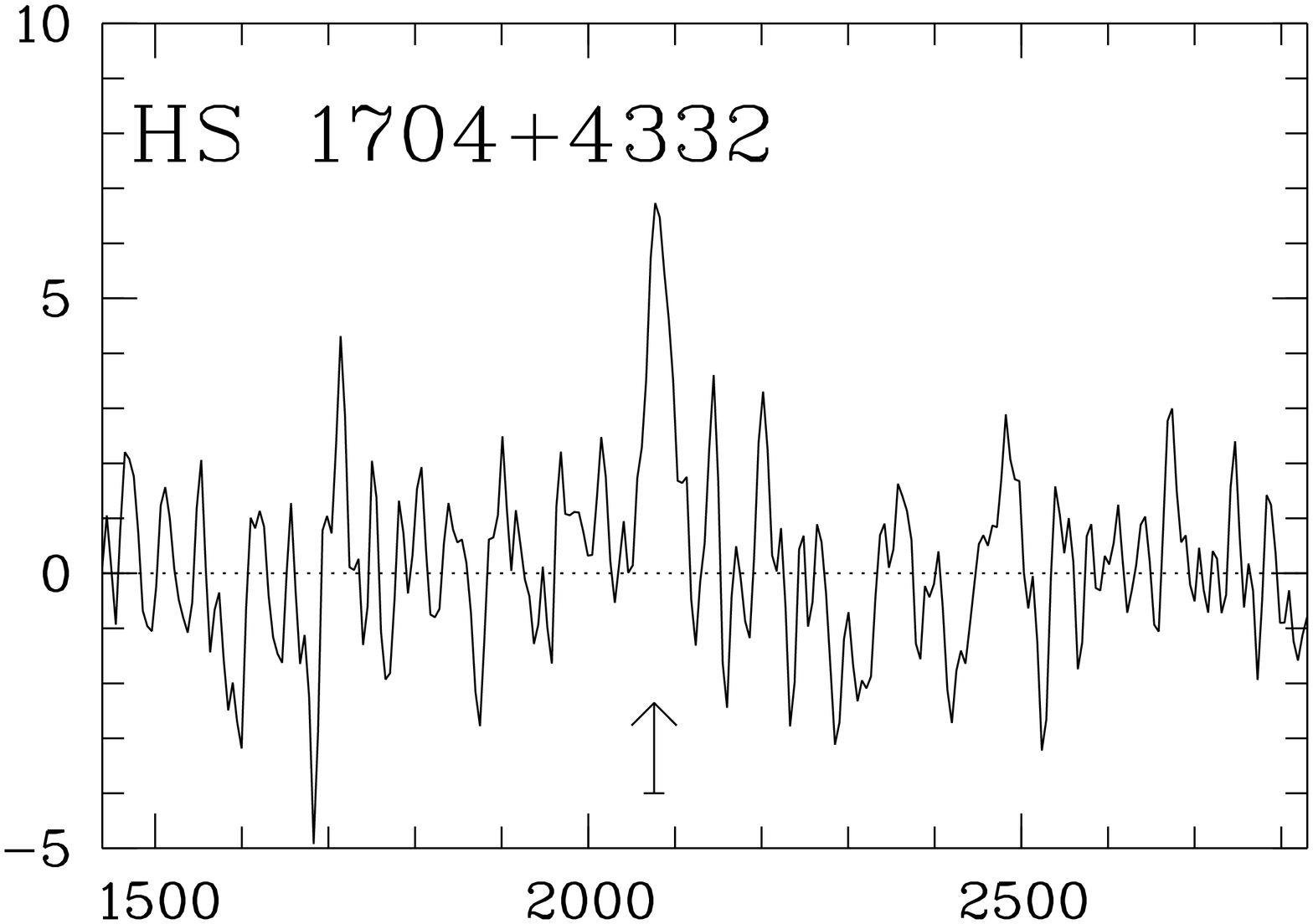} 
 \includegraphics[angle=-90,width=3.8cm, clip=]{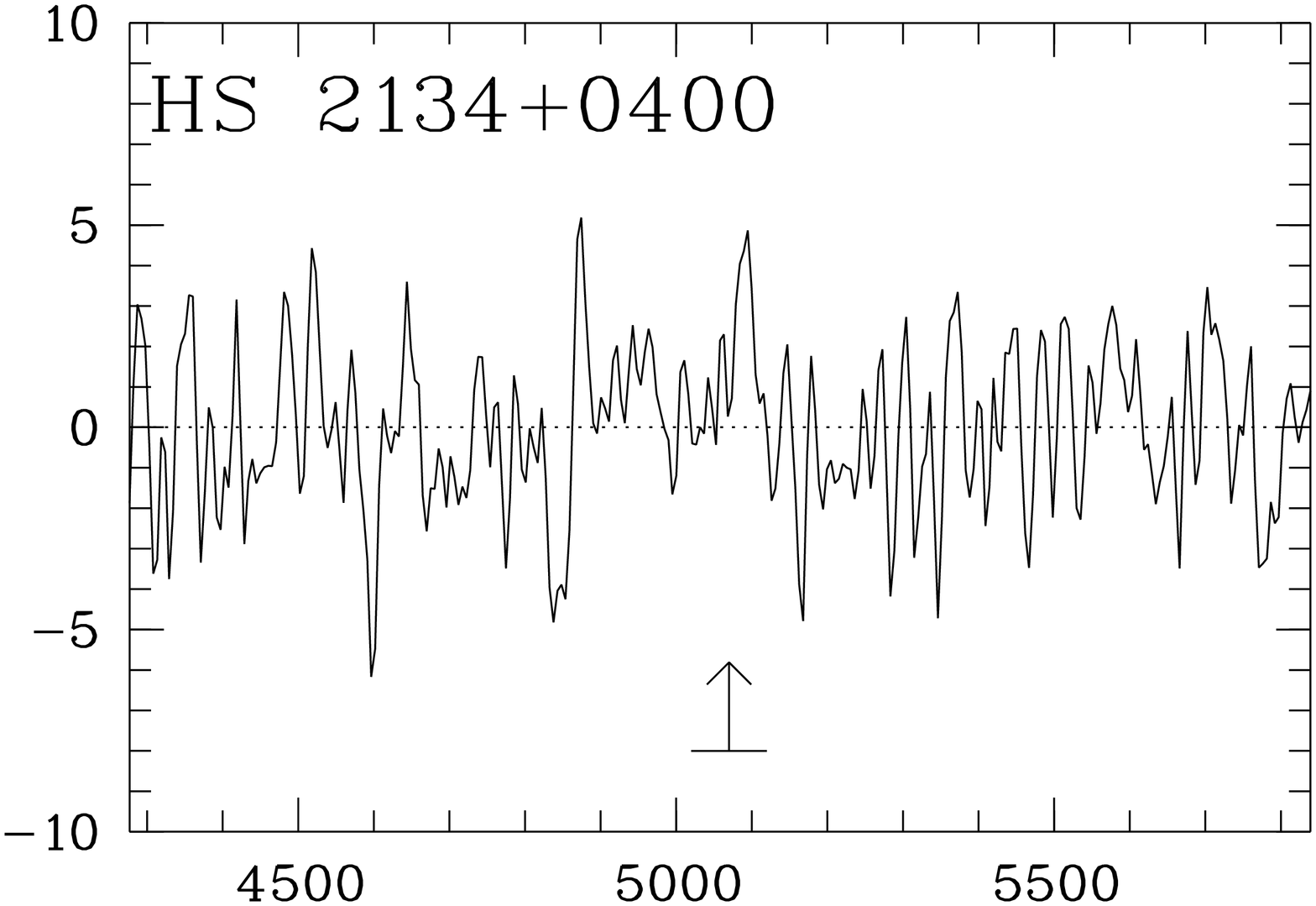} 
 \includegraphics[angle=-90,width=3.8cm, clip=]{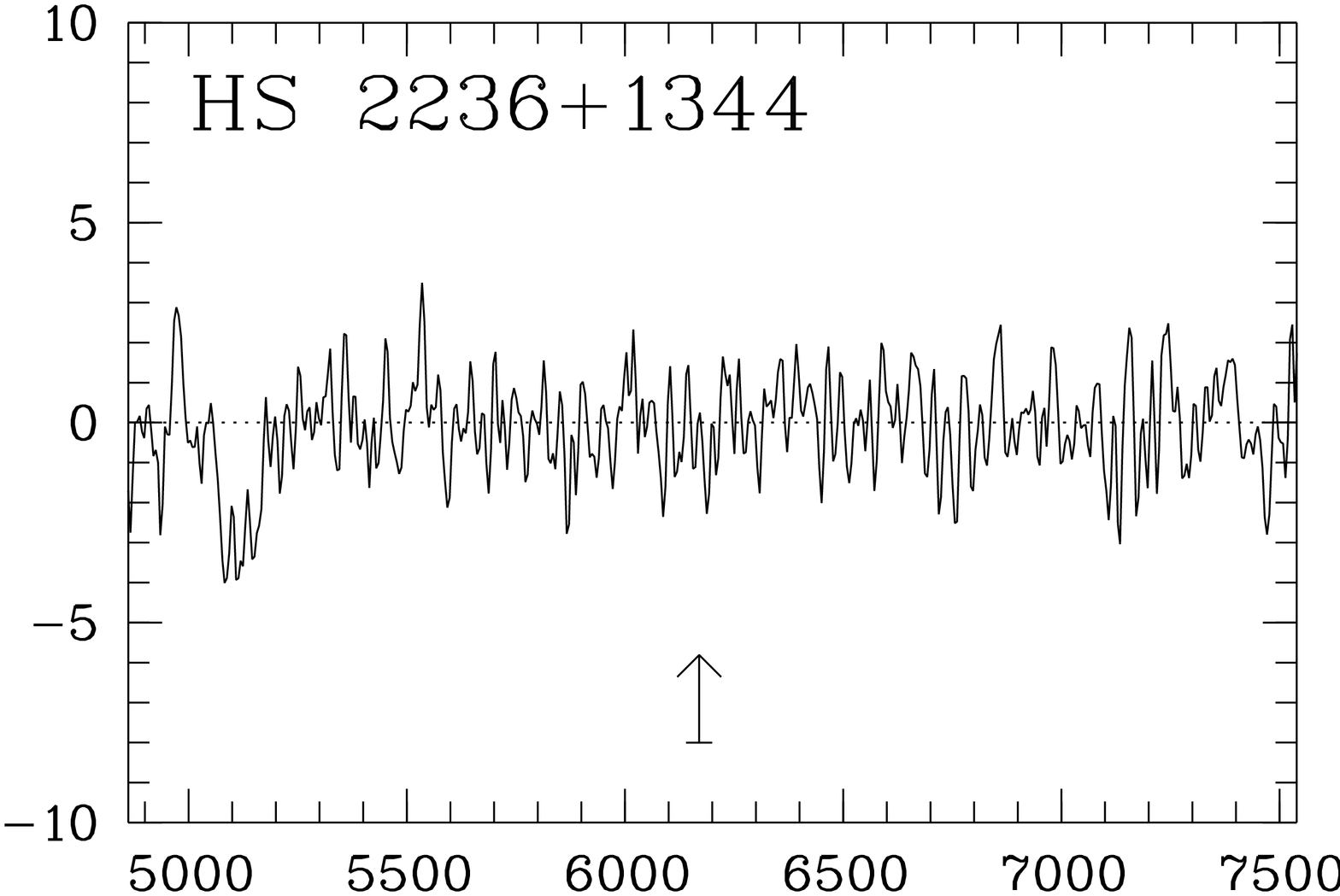} 
   \caption{The NRT \ion{H}{i} profiles S$_{\nu}$ (in mJy) vs. V$_{\rm hel}$
   (\kms) of all studied galaxies with the resolution of 10.5~\kms.
   The vertical arrows at the bottom with the horizontal bars show the
   optical velocity and its $\pm$1$\sigma$\ uncertainty
   for weak detections or non-detected objects.
}
	 \label{Profiles1}
 \end{figure*}

\subsection{HS 0846+3522}

This galaxy is detected near the correct optical velocity of 2177~\kms\
with a S/N ratio of only 3.4. Its velocity in the discovery paper
(Pustilnik et al. \cite{HSS-2}) is mistaken due to a misprint (given as
2481~\kms\ instead of the correct value of 2184~\kms).
Parameters presented in Table \ref{t:HI} have uncertainties of $\sim$50\%.
HS 0846+3522 has a disturbed morphology on the DSS-2 images,
probably due to interaction with an irregular galaxy at 50\arcsec\ NEE
($\sim$7 kpc in projection, $B$-mag $\sim$19.6 from SDSS).
Its redshift is unknown.

\subsection{KISSB 23 = KUG 0937+298}

This galaxy is situated at $\sim$6 Mpc, in the inner rim region of the
nearby  Lynx-Cancer void described in Pustilnik et al. (\cite{SAO0822}).
Two more XMD galaxies are found within the same void (Pustilnik et al.
\cite{SAO0822,DDO68}).
While the blue luminosity of KISSB 23 is only a factor of 2.8 higher
than that of AM~0624--261,
this galaxy, with a similar \ion{H}{i} mass has a significantly broader
\ion{H}{i} profile: $W_{50}$ $\sim$70~\kms.  Even for an edge-on case
this translates to an amplitude of $V_{\rm rot}$ (according to
Staveley-Smith et al. \cite{Staveley92})
of about 35~\kms\ in comparison to 8~\kms\ for AM~0624--261.
Despite having similar parameters of L$_{\rm opt}$ and
M(\ion{H}{i}), KISSB~23 is significantly more DM dominated than AM~0624--261.
This is related to the enhanced stability of its ISM to
local intrinsic perturbations. However, this galaxy shows enhanced SFR.
The EW(H$\beta$) of its bright non-central knot of $\sim$30~\AA\  (Lee et
al. \cite{Lee04}), for the adopted metallicity of $z$=0.001 corresponds to
an age of an instantaneous starburst of $\sim$10 Myr (Leitherer et al.
\cite{Starburst99}).
The origin  of such a  non-central SF ``burst'' is unclear.
The possibility of a trigger by an external perturber can be investigated
through \ion{H}{i} mapping of this galaxy.
Its ratio $M$(\ion{H}{i})/$L_{\rm B}$ = 0.98  is rather high for dIrr
galaxies.

\subsection{HS 0940+4025}

This galaxy is marginally detected (S/N ratio of 2.5), so its parameters have
large uncertainties. The object is very compact and elongated, with the total
extent on the DSS-2 $\sim$9\arcsec\ (3 kpc). The SDSS image shows
two very blue knots almost in contact ($\sim$3\arcsec\ separation), very
similar in appearance to HS~2236+1344, for which there
are clear indications of a merger. The same can be attributed to
HS 0940+4025. Thus, more detailed study of this BCG is necessary.
For its rather small profile width ($\sim$50~\kms), this galaxy has one of
the largest $M$(\ion{H}{i}), only a factor of 2.5 lower than that of the
outstanding XMD BCG Tol~65.

\subsection{HS 1013+3809 and UGC~5540}

This BCG presumably comprises a physical pair with a $\sim$2\fm3 brighter
($M_{\rm B}=-17.7$)  Sc galaxy UGC~5540 at 8\arcmin\ to the South and
$\sim$0\farcm5 to the West (in total $\sim$50 kpc in projection).
The latter was observed in the \ion{H}{i} line with the 91\,m Green Bank
(hereafter GB) telescope and the NRT (Haynes \& Giovanelli \cite{HG91};
Theureau et al. \cite{KLUN7}).
Taking into account the confusion of the two galaxies in the NRT beam,
some useful estimates of the integrated \ion{H}{i} flux of HS 1013+3809 can
be
obtained. If we account for flux attenuation due to an offset of UGC~5540 from
the NRT pointing, we can estimate from the observed flux the residual that
should be attributed to the \ion{H}{i} emission related to HS~1013+3809. In
particular, the value of its \ion{H}{i} mass  can be derived.

The two cited measurements of UGC~5540 ($F$(\ion{H}{i},GB)= 5.37$\pm$0.56 and
$F$(\ion{H}{i},NRT) = 6.20$\pm$1.20 Jy~\kms)
 are not significantly different.
However, the 0.83 Jy~\kms\ higher value of the NRT flux, in comparison to
that measured at GB, may indicate some contribution from HS~1013+3809 to
the pointing toward UGC~5540. Due to the narrower beam for GB, this is not the
case for $F$(\ion{H}{i},GB).
Therefore, for further analysis we accept that the correct $F$(\ion{H}{i}) of
UGC~5540 is given by GB data. Taking into account the attenuation factor
(of 0.655) due to the pointing offset between the two galaxies, the expected
contribution from UGC 5540 at this position is $F$(\ion{H}{i}) =
3.52$\pm$0.37~Jy~\kms.  $F$(\ion{H}{i}) =  3.23$\pm$0.34~Jy~\kms\ when
the horizontal NRT beam resolution effect on the extended
\ion{H}{i} emission of UGC 5540 is applied.
Subtracting this expected flux from the measured value of
4.74$\pm$0.19~Jy~\kms, we estimate the flux for HS 1013+3809
of $F$(\ion{H}{i})=1.51$\pm$0.39~Jy~\kms. The derived value is consistent
with the NRT measured flux of UGC 5540. Indeed, the contribution of
HS 1013+3809 at the NRT pointing to UGC 5540 will be 0.99$\pm$0.25, and
after addition with $F$(\ion{H}{i})=5.37$\pm$0.56 for UGC5540 from GB
data results in 6.36$\pm$0.61, very close to the NRT measured value of 6.20
Jy~\kms. The estimated flux of HS 1013+3809 comprises only 32\% of the
measured \ion{H}{i} flux. To resolve this galaxy pair in \ion{H}{i} requires
mapping with a beam-width of $\sim$1\arcmin.

\subsection{HS 1033+4757}

This very compact and low luminosity galaxy has a bright knot that is somewhat
displaced from the center of the main body. The outer morphology is 
irregular with many filaments. Its \ion{H}{i} profile is rather broad and
asymmetric, suggesting possible confusion. We have checked all sufficiently
bright galaxies falling within the main NRT beam. All those for which
there are SDSS spectra appeared as background objects.

\footnotesize{\bf {
\begin{table*}
\caption[]{\ion{H}{i} observed and derived parameters of the sample and additional (below the line) galaxies}
\label{t:HI}
\begin{tabular}{lrrrrrrrr} \hline \hline
\multicolumn{1}{c}{IAU name }&
\multicolumn{1}{c}{V(HI)} &  \multicolumn{1}{c}{Dist.$^1$} &
\multicolumn{1}{c}{W$_{50}$} & \multicolumn{1}{c}{W$_{20}$} &
\multicolumn{1}{c}{F$_{\rm H}$} & \multicolumn{1}{c}{F$_{\rm c}$} &
\multicolumn{1}{c}{~~~~~~~~Log(M(HI)} & \multicolumn{1}{c}{M(HI)/L$_{\rm B}$} \\
  & \kms & Mpc & \kms & \kms & Jy~\kms & Jy~\kms & M$_{\odot}$~~~~~~~~~ & M$_{\odot}$/L$_{\odot}$~~~~~~ \\
~~~~~(1) &(2)~~~ &(3)  &(4)~~~ & (5)~ & (6)~~~  & (7)~~~ & (8)~~~~~~~~~~ & (9)~~~~~~~~     \\ \hline \\
0017$+$1055   & 5630$\pm$30   &  81.3&   50$\ddagger$~~~~ &  ... & $<$0.20       &               &$<$8.29~~~~~~~~      &$<$0.39~~~~~~  \\  
0122$+$0743   & 2899$\pm$\ 5  &  42.3&   50$\pm$\ 6 & 123$\pm$16 & 6.70$\pm$0.26 & 7.60$\pm$0.30    &9.47~~~~~~~~        & 1.48~~~~~~  \\  
J0133$+$1342  & 2580$\pm$\ 4  &  38.1&   33$\pm$\ 7 &  39$\pm$11 & 0.10$\pm$0.05 &                  &7.54~~~~~~~~        & 0.23~~~~~~  \\  
J0205$-$0949  & 1885$\pm$\ 1  &  26.8&  112$\pm$\ 1 & 133$\pm$\ 2&11.07$\pm$0.19 &12.19$\pm$0.22    &9.35~~~~~~~~        & 2.57~~~~~~  \\  
0624$-$261    &  491$\pm$\ 1  &   5.0&   27$\pm$\ 1 &  42$\pm$\ 2& 2.69$\pm$0.08 & 2.79$\pm$0.08    &7.22~~~~~~~~        & 0.85~~~~~~  \\  
0846$+$3522   & 2169$\pm$\ 3  &  29.6&   33$\pm$\ 6 &  40$\pm$10 & 0.10$\pm$0.03 &                  &7.32~~~~~~~~        & 0.24~~~~~~  \\  
0937$+$2949   &  505$\pm$\ 2  &   6.0&   77$\pm$10  &  98$\pm$15 & 1.95$\pm$0.24 & 2.05$\pm$0.25    &7.24~~~~~~~~        & 0.98~~~~~~  \\  
0940$+$4025   & 5353$\pm$\ 8  &  74.2&   40$\pm$16  &  53$\pm$24 & 0.20$\pm$0.08 &                  &8.42~~~~~~~~        & 0.70~~~~~~  \\  
1013$+$3809   & 1169$\pm$\ 4  &  15.8&   86$\pm$\ 7 & 130$\pm$12 & 4.74$\pm$0.19 & 1.51$\pm$0.39    &7.90~~~~~~~~        & 0.57~~~~~~  \\  
1033$+$4757   & 1541$\pm$\ 9  &  21.8&   86$\pm$\ 7 & 144$\pm$29 & 1.32$\pm$0.15 &                  &8.02~~~~~~~~        & 2.32~~~~~~  \\  
1059$+$3934   & 3019$\pm$\ 3  &  41.9&   59$\pm$\ 6 &  91$\pm$\ 9& 1.39$\pm$0.06 &                  &8.60~~~~~~~~        & 1.54~~~~~~  \\  
J1105$+$6022  & 1333$\pm$\ 3  &  19.9&   48$\pm$\ 6 &  74$\pm$10 & 2.40$\pm$0.17 & 2.48$\pm$0.18    &8.36~~~~~~~~        & 1.62~~~~~~  \\  %
J1121$+$0324  & 1171$\pm$\ 3  &  17.0&   89$\pm$\ 6 & 112$\pm$10 & 2.49$\pm$0.15 & 2.67$\pm$0.16    &8.29~~~~~~~~        & 1.94~~~~~~  \\  
J1201$+$0211$*$& 974$\pm$\ 3  &  17.0&   29$\pm$\ 7 &  53$\pm$10 & 0.96$\pm$0.09 &                  &7.82~~~~~~~~        & 1.92~~~~~~  \\  
1159$+$545    & 3560$\pm$15   &  50.3&   ...        &  ...       &$<$0.10        &               &$<$7.72~~~~~~~~      &$<$0.43~~~~~~  \\  
J1215$+$5223  &  158$\pm$\ 1  &   4.2&   27$\pm$\ 1 &  43$\pm$\ 2& 5.14$\pm$0.14 & 5.24$\pm$0.16    &7.34~~~~~~~~        & 1.01~~~~~~  \\  
1214$-$277    & 7795$\pm$50   & 105.8&   ...        &  ...       &       $<$0.10 &               &$<$8.42~~~~~~~~      &$<$0.28~~~~~~  \\  
1223$-$359    & 2790$\pm$\ 3  &  36.3&   40$\pm$\ 6 &  56$\pm$\ 9& 2.13$\pm$0.24 &                  &8.82~~~~~~~~        & 2.70~~~~~~  \\  
1442$+$4250   &  647$\pm$\ 1  &  10.6&   85$\pm$\ 2 &  99$\pm$\ 2& 6.37$\pm$0.14 & 7.05$\pm$0.16    &8.24~~~~~~~~        & 1.77~~~~~~  \\  
1704$+$4332   & 2082$\pm$\ 8  &  31.9&   33$\pm$15  &  59$\pm$24 & 0.24$\pm$0.05 &                  &7.76~~~~~~~~        & 0.75~~~~~~  \\  
2134$+$0400   & 5090$\pm$\ 4  &  73.9&   25$\pm$\ 9 &  33$\pm$14 & 0.12$\pm$0.05 &                  &8.19~~~~~~~~        & 0.76~~~~~~  \\  
2236$+$1344   & 6160$\pm$20   &  89.1&   50$\ddagger$~~~~ &  ... &$<$0.15        &               &$<$8.35~~~~~~~~      &$<$0.25~~~~~~  \\  \hline 
HI0624$-$2614 &  188$\pm$\ 2  &   0.02&   37$\pm$\ 5 &  58$\pm$\ 7& 0.77$\pm$0.06 &               &$>$1.86~~~~~~~~        &     ~~~~~~  \\ 
1059$+$395    & 3023$\pm$\ 5  &  41.9&   62$\pm$10  & 101$\pm$16 & 1.38$\pm$0.09 &                  &8.59~~~~~~~~        & 0.86~~~~~~  \\ 
1102$+$6038   & 1265$\pm$\ 8  &  19.0&   43$\pm$\ 9 &  70$\pm$15 & 2.30$\pm$0.17 &  2.58$\pm$0.19   &8.34~~~~~~~~        & 1.52~~~~~~  \\ 
J1106$+$6015  & 1252$\pm$\ 8  &  19.0&   57$\pm$\ 9 &  80$\pm$15 & 2.10$\pm$0.18 &                  &8.25~~~~~~~~        & 1.02~~~~~~  \\ 
J1202$+$0215  & 2107$\pm$\ 8  &  27.1&   33$\pm$16  &  76$\pm$24 & 0.83$\pm$0.10 &  0.97$\pm$0.12   &8.23~~~~~~~~        & 4.5:~~~~~~  \\ 
HIJ1218$-$2801& 7392$\pm$\ 3  & 100.2&  117$\pm$\ 6 & 127$\pm$10 & 0.46$\pm$0.06 &                  &9.04~~~~~~~~        &     ~~~~~~  \\ 
1342$+$4210   & 3850$\pm$21   &  54.6&   70$\pm$42  & 120$\pm$67 & 0.40$\pm$0.10 &                  &8.32~~~~~~~~        & 1.21~~~~~~  \\ 
2236$+$136    & 5122$\pm$10   &  75.8&  105$\pm$16  & 120$\pm$25 & 0.30$\pm$0.08 & 0.63$\pm$0.17    &8.91~~~~~~~~        & 0.15~~~~~~  \\ 
\hline\hline \\
\multicolumn{8}{l}{$^{1}$ -- Distances adopted similar to Karachentsev et al. (2004) with H$_{0}$ = 72~\kms~Mpc$^{-1}$. }  \\
\multicolumn{8}{l}{$\ddagger$ -- Adopted from the total ionized gas velocity amplitude }  \\
\multicolumn{8}{l}{$*$ -- Adopted to be the distance of Virgo cluster} \\
\multicolumn{8}{l}{{\bf 0017$+$1055} -- $\sigma_{\rm noise}$ = 1.8 mJy. For the F(HI) upper limit the value $\Delta$V=50~\kms\ is used from H$\alpha$ data. }                       \\
\multicolumn{8}{l}{{\bf J0205$-$0949} -- Correction factor of 1.10, calculated as for Sc galaxy NGC 5540, see Sect. \ref{J0205}.  }  \\
\multicolumn{8}{l}{{\bf 0624--2614} --  Distance from Karachentsev et al. (2004) }  \\
\multicolumn{8}{l}{{\bf 0937+2949} --  Distance is estimated as in Karachentsev et al. catalog (2004), where this galaxy is absent. }\\
\multicolumn{8}{l}{{\bf 1013+3809} --  Resolved confusion with Sc galaxy of $\sim$2 mag. brighter, at 8\arcmin\ to the S, see text. } \\
\multicolumn{8}{l}{{\bf 1059+3934} --  Resolved confusion with KUG 1059+395 at 1.9\arcmin, see text.}      \\
\multicolumn{8}{l}{{\bf J1105+6022} -- In a group of three dwarfs, see text. } \\
\multicolumn{8}{l}{{\bf 1159$+$545} -- $\sigma_{\rm noise}$ = 1.0 mJy.  }                       \\
\multicolumn{8}{l}{{\bf J1215+5223} -- Correction factor of 1.02, accounting for \ion{H}{i} structure from GMRT (Begum et al. \cite{Begum06}) }      \\
\multicolumn{8}{l}{{\bf 1214--277} --  $\sigma_{\rm noise}$=1.0 mJy.  FLASH J121809.29$-$275219.5 is probably a member of the same group. }                                         \\
\multicolumn{8}{l}{{\bf 1223--359} --   A galaxy at 2\farcm1 to the East and 4\arcmin to the South.}       \\
\multicolumn{8}{l}{{\bf 2236$+$1344} -- $\sigma_{\rm noise}$=1.2 mJy. For the F(HI) upper limit the value $\Delta$V=50~\kms\ is used from H$\alpha$ data. }  \\
\multicolumn{8}{l}{{\bf HI0624$-$2614} -- HVC? Assumed $D=$ 20 kpc. Unknown correction for angular size/offset.}  \\
\multicolumn{8}{l}{{\bf 1059$+$395} --  Observed to resolve confusion for HS 1059+3934. See text. }  \\
\multicolumn{8}{l}{{\bf 1102$+$6038} -- MCG+10-16-056, observed to resolve confusion for SDSS J1105+6022. }  \\
\multicolumn{8}{l}{{\bf J1202$+$0215} --  LSB/dI galaxy near J1201+0211 OFF-source position. Flux is corrected for offset attenuation.}  \\
\multicolumn{8}{l}{{\bf HIJ1218$-$2801} -- A source near OFF-source position J121830.74$-$280125. No NED candidates.  }  \\
\multicolumn{8}{l}{{\bf 2236$+$136} -- In case this is KUG 2236+136 seen in the OFF-source position, the offset attenuation is}   \\
\multicolumn{8}{l}{~~~~~~~~ of $\sim$2.1. However, this can be a blend since its velocities in UZC (5188~\kms) and SDSS (5224~\kms) } \\
\multicolumn{8}{l}{~~~~~~~~ catalogs differ significantly from the value found here. }  \\
\end{tabular} 
\end{table*}
     }
 }
\normalsize

\subsection{HS 1059$+$3934 and KUG 1059$+$395}

The measured \ion{H}{i} flux for the NRT pointing at the position of
HS~1059+3934  is 1.39 Jy~\kms.
Its \ion{H}{i} profile is quite narrow ($W_{50}$=59 \kms), indicating a
small rotation velocity.
The inclination correction ($i$=62\degr, $sin~i$=0.88) is not large, at
least as it can be estimated from the DSS-2 image, where the axial ratio
$b/a \approx$0.5.
The nearest galaxy, which could potentially affect the \ion{H}{i} signal
of HS~1059+3934, is KUG 1059+395, an irregular galaxy with a
bright central knot,  1\fm2  brighter than the target galaxy,
at 1\farcm9 to the West (25 kpc in projection).
We obtained the optical spectrum of KUG 1059$+$395  with the SAO 6m
telescope and found this to be a starburst galaxy with radial
velocity close to that of HS~1059+3934 (Pustilnik et al. \cite{BCG_abun}).

To resolve the potential confusion, we carried out NRT observations
of KUG 1059+395. The observed \ion{H}{i} profile is centered at
a velocity very close to that for HS 1059+3934. A small
extra feature is seen at the high velocity end. The derived integrated flux
(without the extra feature) is 1.38 Jy~\kms. Having two independent
measurements and a known attenuation factor due to offset positions (0.475)
for both pointings, we solved this system in several iterations.
The total flux is 0.95 Jy~\kms\  for HS 1059+3934 and 0.93 Jy~\kms\
for KUG 1059+395. The extra feature detected for KUG 1059+395, with the total
flux of 0.3 Jy~\kms, is probably an artifact.
The recent GMRT mapping of these galaxies (in preparation) shows
disturbed \ion{H}{i} morphology in both galaxies, implying strong
interaction in the system.

\subsection{SDSS J1105+6022 and its neighbors}

There is some \ion{H}{i} signal in the off-source position, which looks
as a negative amplitude profile centered at $V_{\rm hel} \sim$1260 \kms.
We have checked possible candidate galaxies, and found that the most probable
is MCG+10-16-062, a galaxy of $B$=16.3, which is
at $\Delta \alpha = 7^{s}.0$  (52\arcsec) and $\Delta \delta$=6\farcm7 from
the off-source pointing. Its radial velocity in SDSS is given
as 1273$\pm$3~\kms. Its offset results in the attenuation
factor of 0.75. This galaxy is situated at 9\farcm4 SE from our target
object.

Another potential source of confusion is a disturbed DIG/LSB galaxy of a
comparable size and total magnitude, MCG+10-16-056 (m$_{\rm B} \approx$16.5),
which is only at
2\farcm4 west from the target XMD galaxy. Its redshift is unknown.
Since this can also contribute to the measured \ion{H}{i} flux, we
conducted an NRT observation with  pointing to MCG+10-16-056.
It was detected at a velocity of 1265 \kms, which is very close to
that of MCG+10-16-062. The contribution of the latter in
the off-source position for this observation was also significant.
Having these 3 results, we solve
this system by iterating the fluxes for all three galaxies. This
method converges within the observational errors.
The derived parameters of these galaxies are given in Table \ref{t:HI}.
The XMD galaxy and these two new \ion{H}{i} detected dwarfs (at the projection
distances of 14 and 55 kpc) presumably form an isolated `dwarf' group.
The preliminary results  of the GMRT mapping (in preparation)
reveal  \ion{H}{i}-line emission from all three described above
galaxies, with \ion{H}{i} morphology of SDSS J1105+6022 and MCG+10-16-056
typical of a merging pair.

\subsection{SDSS J1121$+$0324 = NGC~3640A}

This galaxy is well detected with a S/N ratio of $\sim$10. Its
profile, resembling a  double-horn   one, is consistent with its optical
appearance as an inclined disk. This disk has two bright regions
displaced from the center and rather disturbed outer isophotes.
Hibbard \& Sansom (\cite{Hibbard03}) show this galaxy (called NGC 3640A)
to be a probable companion of elliptical galaxy NGC 3640 (at a projected
distance of $\sim$70 kpc), which belongs to galaxy group LGG 233. The
\ion{H}{i}-line VLA map of this
region shows no confusing galaxies. Accordingly, its integrated parameters,
derived in that work, are consistent with those obtained on the NRT data.

The $B$-magnitude of 17\fm91 from Kniazev et al. (\cite{SDSS}) relates only
to emission from the bright \ion{H}{ii} region.
Our independent estimate of its magnitude (from SDSS) of
$B$=17\fm88 is consistent with theirs. The $B$-magnitude for the rest of the
galaxy (also from SDSS photometry) is 17\fm18. Summing up the light of
both objects results in $B_{\rm tot}$=16\fm92.

\subsection{SDSS J1201+0211}
\label{J1201}

This galaxy with rather disturbed morphology is a probable merger with the
fainter ($B \sim$20.4) companion at 17\arcsec\  NW (1.4 kpc in projection).
It lies at the
southern edge of the Virgo Cluster. No known galaxies in NED with the radial
velocities within our velocity  range are expected to be in the NRT beam.

A significant signal, corresponding to an object near the
off-source position, is detected at V$_{\rm hel}$ of $\sim$2110 \kms.
The latter falls within the range of the Virgo cluster
members in the direction close to its center. However, for the distant
southern periphery this velocity is too high and this \ion{H}{i}
object is far  behind the Virgo cluster.
Its W$_{\rm 50}$ is $\sim$30~\kms, suggesting very slow rotation or large
inclination.
There are two candidate faint optical galaxies within the NRT beam near this
position. The first one is an edge-on disk, at 100\arcsec\ N,
25\arcsec\ W, and the second is a face-on LSBG, at 288\arcsec\ N,
25\arcsec\ W.
Considering the very narrow \ion{H}{i} profile, the `face-on' LSBG
galaxy is a more probable counterpart. However, its $B \sim$18.8
(from SDSS) implies a very high ratio M(HI)/L$_{\rm B} \sim$4.5.
This requires further study.

\subsection{SDSS J1215$+$5233 = CGCG 269$-$049}

This galaxy belongs to the Canis Venatici (CnV) I galaxy cloud at a
distance of $\sim$4 Mpc (e.g., Karachentsev et al. \cite{Kara03}).
Its narrow \ion{H}{i} profile, with W$_{\rm 50} =$25~\kms,
is consistent with its very low optical luminosity. The inclination
correction for its $b/a=$0.3 (Karachentsev et al. \cite{Kara04}) is small,
$\sim$1.03.
Its nearest neighbor galaxy UGC~7298 (at the projected distance of 14.3 kpc)
has a similar radial velocity of 146 \kms. The distance of UGC~7298 of
4.21 Mpc is determined through the tip of the RGB (Karachentsev et
al. \cite{Kara04}). We therefore consider the distance to SDSS J1215$+$5233
to be equal to 4.2 Mpc as well. These galaxies seem to be
a close pair and tracers of the recent/current interaction might be visible
in their \ion{H}{i} morphology. Indeed, the \ion{H}{i} maps presented by
Begum et al. (\cite{Begum06}), give some hint of disturbance in both
galaxies of the neutral gas morphology and its kinematics.

\subsection{Tol 1214$-$277 = Tol~21}

This BCG is the most distant in our sample.
Despite the long integration and the low noise ($\sigma \sim$1~mJy for
the effective velocity resolution of 10.5~\kms), no signal was detected in
the spectrum at or near the optical velocity.
However, we have detected an object near the off-source position
(a negative signal with a  `double-horn' profile) with
W$_{\rm 50} =$117~\kms, at $V_{\rm hel}$ of 7392~\kms.
We have checked known galaxies close to the off-source beam position (72.5 s
in RA, or 16\arcmin\ to the East) and found no appropriate candidates in NED.
Several galaxies around Tol~21 have $V_{\rm hel}$ of 7600--7740~\kms,
implying that Tol~21 probably belongs to a loose group.
If this \ion{H}{i} object is a galaxy with M(HI)/L$_{B}$ $\sim$1, its
measured integrated flux F(HI) of 0.46 Jy~\kms\ corresponds to
$B_{\rm tot} \sim$18. Accounting for a probable offset and a lower value
M(HI)/L$_{B}$, it can be even brighter.

The upper limit for the \ion{H}{i} mass of Tol~21 is derived assuming that its
W$_{\rm 50} <$ 100~\kms\ and the peak value of the \ion{H}{i} profile is lower
than 2 mJy.
The upper limit for the ratio M(\ion{H}{i})/$L_{\rm B}$ is low
($<$0.28). For most of other XMD galaxies this parameter is (significantly)
larger. This implies
that this object belongs to the lower end of the XMD galaxy \ion{H}{i}
mass function.
Probably the neutral gas ``deficiency'' is related to the detectability of
Ly-$\alpha$ emission in this BCG.
This galaxy  is the only one among four XMD BCGs in which
Ly-$\alpha$ emission was detected in the HST observations (Kunth et al.
\cite{Kunth94}; Thuan \& Izotov \cite{TI97}). The very faint galaxy
G2 with close $V_{\rm hel}$ (Izotov et al. \cite{Izotov04}) at a projected
distance of 14.5 kpc is probably a companion of Tol~21.

\subsection{Tol 1223$-$359 = Tol~65}

This galaxy is quite unusual since its \ion{H}{i} profile is rather narrow
for its large \ion{H}{i} mass of 0.66$\times$$10^{9}~M$\sunn.
The inclination correction is difficult to infer since its optical
morphology is strongly disturbed at outer isophotes (Papaderos et al.
\cite{Papa99}).
It resembles  SBS~0335$-$052~E, whose \ion{H}{i} mass is only a factor of
1.2 higher. Their total \ion{H}{i} linewidths are also similar.

Tol~65 is situated at the outskirts of a galaxy group.
There are several galaxies with close radial velocities at angular distances
of 38\arcmin\ to 53\arcmin. 
One irregular galaxy, FLASH J122530.92$-$360714.5 (Kaldare et al.
\cite{FLASH}), with $V_{\rm hel}$ = 2781$\pm$56~\kms\ and $m_{b} =$15.7,
at 7\farcm5 to the NW, could contribute some \ion{H}{i} emission to the
observed
flux. However, due to offset attenuation, its measured flux will be
$\sim$9\%  of the actual value.
Even for M(\ion{H}{i})/L$_{\rm B} \sim$1, this galaxy would
contribute no more than 20\% to the total \ion{H}{i} flux of Tol~65.
The close LSBD G1 (Papaderos et al. \cite{Papa99}) is about 5 magnitudes
fainter and would give only a minor contribution to the observed \ion{H}{i}
flux.

\subsection{HS 1442+4250 = UGC 9497}

This galaxy appears to be an edge-on disk with two prominent off-center
\ion{H}{ii}-regions and several fainter ones (Kniazev et al. \cite{Kniazev98};
Guseva et al. \cite{Guseva03b}).
It is one of the nearest XMD BCG known. Its deep $V,I$ photometry and
long-slit spectroscopy show no evidence of a stellar population older
than 2 Gyr (Guseva et al. \cite{Guseva03b}). Its high S/N ratio \ion{H}{i}
double-horn  profile is rather wide, indicating significant rotation.
Some asymmetry in its \ion{H}{i} profile at the high-velocity edge of
$\sim$10\% of the peak value may indicate a low mass companion, or a galaxy
within the NRT beam. However, a better S/N ratio profile is necessary to
confirm this. There is a fainter red galaxy almost in contact, but its radial
velocity is unknown, so this might be a background object.

\subsection{HS 1704+4332}

This galaxy has quite a narrow profile, consistent with its low luminosity.
The galaxy  is situated in the region of low density  luminous galaxies
($L \gtrsim L_{*}$) with the nearest one being situated at 4.7 h$^{-1}$ Mpc.
The nearest known subluminous Updated Zwicky Catalog (UZC) galaxy
is at 3 h$^{-1}$ Mpc.

\subsection{HS 2134+0400}

This is the most metal-poor BCG and the faintest galaxy in $B$-band in this
subsample. The expected \ion{H}{i} flux is hence also low.
There is a narrow peak (W$_{50}$=25 \kms) at a velocity close to the
optical one, but its amplitude is not very significant, of the order of
3 $\sigma_{\rm noise}$. Its morphology on the deep $B$-band image looks rather
disturbed, with a plume at the W edge that suggests its probable merger
nature (Pustilnik et al. \cite{HS2134}), since no disturbing galaxies are
present in its environment. The nearby small irregular galaxy South of the BCG
is a distant background object.

\subsection{Non-detected XMD galaxies}

Of four undetected XMD galaxies three are the most distant in the sample,
with V$_{\rm hel}$ in the range of $\sim$5600 to 7800~\kms.
For these four galaxies (HS~0017+1055, SBS~1159+545, Tol~21, and
HS~2236+1344), the Notes for Table \ref{t:HI} give estimates
of $\sigma_{\rm noise}$ for the resolution of 10.5~\kms\ in the spectral
regions near their optical velocities.
To derive upper limits of their $M$(\ion{H}{i}), we assumed that their
peak flux densities $F_{\rm peak}$ are lower than the visible peak flux
density near the optical velocity plus 2$\sigma_{\rm noise}$ (in mJy), and
that their profiles are boxcar function with $W_{50}$ less than 100 \kms.
Then their integrated   \ion{H}{i} flux $F$(\ion{H}{i}), is less than
$F_{\rm peak}\times$ $W_{50}$.
For SBS 1159+545 the situation is complicated since close to the optical
velocity we have a faint negative signal, indicating possible confusion
with an object in the `OFF'-source beam. However, we failed to find
a reliable optical counterpart for this negative signal. If the negative
peak near the optical velocity is treated as noise, a 3$\sigma$ upper
limit of the peak
flux density for SBS 1159+545 is $\sim$1~mJy. We present its parameters
based on this value in Table \ref{t:HI}. For $W_{50}$ in the spectra
of  HS~0017+1055 and HS~2236+1344 we used the full amplitude of the
ionized gas velocity as measured in the H$\alpha$-line ($\sim$50 \kms)
from the SAO 6\,m telescope long-slit observations (in preparation).

The ratio $M$(\ion{H}{i})/L$_{\rm B}$ $<$ 0.25 for HS~2236+1344 is one of
the lowest among the XMD BCGs observed in this work. There are two additional
factors that could affect its value. They both are related to its probable
`atypical' nature. As the analysis of its
morphology shows (two bright, almost contacting star-forming knots with
disturbed periphery and plumes)  and the unusual ionized gas kinematics
(Pramskij et al. \cite{Pramskij03}; Pustilnik et al. in preparation), this
object is very likely an advanced merger. The strengths of starbursts in
mergers are the largest in comparison to other cases since the main fraction
of available gas can be involved in SF. This implies that in mergers, the
$M$(\ion{H}{i})/L$_{\rm B}$ ratio is shifted down from its `typical' value.
The second reason is that conditions in the ISM of merging
gas-rich galaxies are favorable for transforming the whole \ion{H}{i}
cloud population to H$_{2}$ (e.g., Elmegreen \cite{Elmegreen93}).

\subsection{Non-XMD galaxies}

The parameters of the observed and detected non-XMD galaxies are presented
at the bottom of Table \ref{t:HI}. We divide them into three categories.

{\it First}: the targets previously considered as XMD galaxies, but later
eliminated from this group due to better precision O/H determination.
The only galaxy in this category is HS 1342+4210, with the
value of 12+$\log$(O/H)=7.89. This is probably interacting with a
fainter companion at 13\arcsec\ SEE, as seen on the DSS-2 image.

{\it Second}: the targets observed to resolve the potential or
certain confusion, either in the `ON'-source, or in the `OFF'-source beam.
They include KUG 1059+395, MCG+10-15-056 and MCG+10-15-062.

{\it Third}: new \ion{H}{i} sources  which appeared    occasionally
in the `OFF' beam in the range of \,$\pm$1300~\kms\ near the
velocity of the target XMD galaxy. One of them, near the position of HS
2236+1344, could be tentatively identified (based on the information from
NED) with the cataloged galaxy KUG 2236+136. Another faint galaxy,
J1202+0215, detected
near the position of SDSS J1201+0211 (see Sect. \ref{J1201}) is not cataloged.
For the faint source HI J1218$-$2801 near Tol~21 we did not find any optical
counterpart.

\section{Discussion and conclusions}
\label{discussion}

As mentioned in the introduction, four evolution scenarios can possibly
explain the existence of  XMD galaxies. For old galaxies with very low SFR
(kind of LSBGs) and truly young galaxies for which one expects the closed-box
to be valid,  the small range of O/H in our
sample corresponds, according to  estimates, e.g., of Pilyugin \&
Ferrini (\cite{PF00}) to a gas mass fraction $\mu >$ 0.9.
However, if the low metallicity of a galaxy ISM is acquired due
to exchange with the intergalactic medium, the gas mass-fraction can be
significantly lower, e.g., M$_{\rm HI}$ $\lesssim$ M$_{\rm stars}$.
The estimate of \ion{H}{i}  mass from the integrated \ion{H}{i} flux
is quite straightforward. However, the estimate of the total stellar mass
in star-bursting galaxies requires the separation of the flux from the
young population and surrounding \ion{H}{ii} regions that often
dominate hiding an old,  underlying component.
Thus, to address the issue of the gas-mass fraction, we need
good surface photometry of the studied objects. We just mention that
in addition to well-known very gas-rich XMD galaxies I~Zw~18 and
SBS~0335--052, the \ion{H}{i} data for Tol~65 from this paper, along
with the surface photometry from Papaderos et al. (\cite{Papa99}), indicate
that this galaxy belongs to the same class.
We postpone the related analysis based on photometric data
to a forthcoming paper.
Below we discuss the large variance of \ion{H}{i} parameters of
the XMD galaxies and new evidence for the importance of interactions
to trigger starbursts in the XMD BCG sample.

\subsection{Variance of \ion{H}{i} and optical properties}

\begin{figure}
   \vspace*{0.5cm}
   \centering
 \includegraphics[angle=-90,width=7.0cm,clip=]{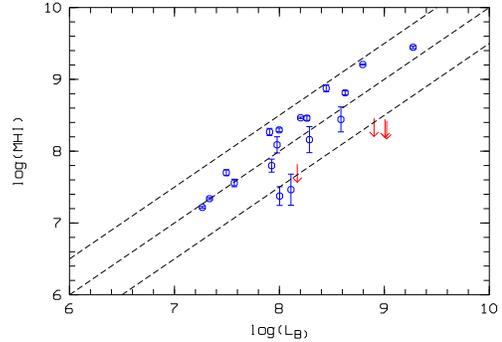}
   \caption{
       Relation between M(\ion{H}{i}) and L$_{\rm B}$ for the XMD galaxies
	   from this paper. Middle dashed line corresponds to positions of
       objects with M(\ion{H}{i})/L$_{\rm B}$=1.0, while the upper and
       lower lines for objects with this ratio equal 3 and 1/3, respectively.
       The arrows show the upper limits on M(\ion{H}{i}).  }
	 \label{MHI_vs_LB}
 \end{figure}

We briefly summarize the data in Tables \ref{t:Param} and \ref{t:HI}.
The total range of oxygen abundances in the sample is small, 0.23 dex, or
a factor of 1.7. However, the total range of $B$-band luminosities
in the studied XMD galaxies is more than two orders of magnitude
(M$_{\rm B}^{0}$ from --12.5 to --17.7, or $\log$(L$_{\rm B}$/L\sunn) = 7.19
to 9.27).
The range of \ion{H}{i}  mass in the sample XMD galaxies is comparable
to the optical luminosity range: $\log$(M(\ion{H}{i})/M\sunn)=7.22 to 9.47.
The latter implies that for galaxies in the very-low metallicity regime
this is not tightly connected with their other global
parameters.  For the blue luminosity, this can be partly understood
since the studied galaxy sample includes objects of various types
- LSBDs, DIGs, BCGs, and even merging pairs. The strong starbursts in BCGs
can brighten the blue luminosity of a progenitor galaxy by a factor of
$\sim$10 or more (e.g.,
Kr\"uger \cite{Kruger92}), depending on the starburst strength, its age, and
the preburst evolutionary state. This brightening, however, could
account only for a part of this large range of L$_{\rm B}$.
 Figure \ref{MHI_vs_LB} shows that the L$_{\rm B}$ and M(\ion{H}{i})
of our XMD galaxy sample are correlated, and have a mean
M(\ion{H}{i})/L$_{\rm B}$ of 1.15. The scatter of the data is smaller
than a factor of $\sim$2.4. However, five of the sample galaxies
show values of M(\ion{H}{i})/L$_{\rm B}$ that are below the mean by a
factor of 3--5.  The latter can indicate evolution with significant
metal loss, as for the nearby old XMD DIGs Sextans A and B, with
M(\ion{H}{i})/L$_{\rm B} < 0.2$.

\subsection{Interactions as a SF trigger in XMD BCGs}

The majority of known XMD galaxies are BCGs,  low-mass galaxies
with sufficiently strong SF activity. One of the important questions
on XMD BCGs is the nature of the triggering mechanisms for their SF bursts.
Both external and intrinsic mechanisms are proposed to be responsible for the
starbursts observed in BCGs in general, e.g., Salzer \& Norton (2000) argue
that BCG starbursts are due to an intrinsic trigger in some specific
BCG progenitors -- gas-rich galaxies with  enhanced gas concentration
index. However, if some XMD galaxies are really young, the external
trigger due to gravitational interactions with other galaxies appears
more natural,  otherwise it is not clear why they did not start SF much
earlier. The interaction trigger seems to be sufficiently important
for starbursts in BCGs in general (e.g.,
Taylor et al. \cite{Taylor93}, \cite{Taylor95}; Pustilnik et al.
\cite{Pustilnik01b}). Therefore, it is natural to check whether sufficiently
close companions/neighbors of various masses exist in the vicinity of our
target galaxies.

The majority of galaxies in the studied sample appear to have various
indications of stronger or weaker interaction. First of all there are
direct indications of close companions (both  in terms of projected
distance and  radial velocity).
This relates to HS 1059+3934 and KUG 1059+395 at 25 kpc, HS 1013+3809
and UGC~5540 at 50 kpc, and SDSS J1105+6022 and MCG 10-16-056
at 15 kpc. SDSS J1121+0324 is a probable companion of E3 galaxy NGC 3640
at 74 kpc and a member of the galaxy group LGG 233.
SDSS J1215+5223 is paired with UGC 7298 at 14 kpc. Tol~21 has a
tiny companion at 14 kpc (Izotov et al. \cite{Izotov04}).
One more XMD BCG HS 0822+3542 has a companion LSBG at 11 kpc
(Pustilnik et al. \cite{SAO0822}; Chengalur et al. \cite{GMRT}).
Several other XMD galaxies HS 0122+0743, HS 2236+1344,  and HS 0837+4717
(Pramskij et al. \cite{Pramskij03}; Pustilnik et al. \cite{HS0837}) show
clear evidence for various stages of a merger from both their optical
morphology and  ionized gas kinematics.

Other XMD BCGs show only indications for interaction-induced SF activity
based on their morphology. In particular, HS 0846+3522 and SDSS
J1201+0211 have a disturbed external morphology and
fainter galaxies in their vicinity, suggesting ongoing merging.
HS 2134+0400 has a disturbed morphology of the outer parts and a plume on the
western edge. Another indication of interaction is an asymmetric
\ion{H}{i} profile, which is visible in HS 1033+4757.

In summary, we note that the majority of XMD BCGs are either certainly or
probably interacting. Our
\ion{H}{i} studies gave several new indications for the importance of
tidal trigger to SF activity. However, more detailed studies of
\ion{H}{i} morphology and kinematics, as well as those of the ionized gas
in these galaxies, are necessary to unambiguously support this 
conclusion.

\subsection{Probable new high-velocity cloud}
\label{HVC}

A well-detected \ion{H}{i} source in the `ON' beam in the direction of
AM~0624--261, at V$_{\rm hel}$ = 188~\kms\ can be seen in Fig.
\ref{Profiles1}.
Its parameters are given in Table \ref{t:HI} under the name HI~0624--2614.
No candidate galaxies within the NRT beam that could be  counterparts
for this \ion{H}{i} source are found.

Therefore, the most probable interpretation of this feature is the emission
of a high-velocity \ion{H}{i} cloud (HVC) near the Milky Way. In the
HVC name system, it would be HVC 234.3-16.8+208 (which means galactic
longitude and latitude in degrees and the local standard of rest velocity in
\kms). The nearest known HVCs to this position  are HVC 234.6-17.2+175,
HVC 233.6-17.2+175, and HVC 233.8-18.7+212 from the HIPASS HVC catalog of
Putman et al. (\cite{Putman02}), at  angular distances of
0.5$^{\circ}$, 0.8$^{\circ}$, and 2.0$^{\circ}$, respectively.
The FWHMs of their \ion{H}{i} profiles are 33--34~\kms\ in comparison to
the value of 37$\pm$5~\kms\ measured for HI~0624--2614.
These nearby HVCs have axial ratios of $a/b$ $\sim$1.5, with the major axis
FWHM of 0.8$\degr$ to 2.4$\degr$ and peak brightness temperatures
T$_{\rm b}$ of 0.06 to 0.23~K. Their total \ion{H}{i} fluxes vary in the
range of 9 to 97 Jy~\kms. The NRT beam with FWHM =
0.37$\degr \times$0.06$\degr$ is rather small for such HVCs and due to
resolution will pick up only a fraction of the total \ion{H}{i} flux.
The flux loss can
vary from a factor of $\sim$10 to several hundred. Therefore, if this source
is a part of a HVC, the measured flux of this object 0.77 Jy~\kms\
matches the known range well. Note, that for an extended source with
a characteristic size of $\gtrsim$10\arcmin, a large reduction of the
measured signal will occur due to subtraction of the source signal
in the `OFF' beam (here at 13\farcm5).
Thus, all parameters of HI~0624--2614 are consistent with the hypothesis
that it is a portion of a HVC similar to several such nearby  objects.

\subsection{Conclusions}

\begin{enumerate}
\item We have measured the integrated \ion{H}{i} parameters for 18 extremely
    metal-deficient (12+$\log$(O/H) $\le$7.65)
    dwarf galaxies, including 15 measured for the first time. For four more
    XMD galaxies we obtained useful upper limits on their \ion{H}{i} mass
    and ratio $M$(\ion{H}{i})/$L_{\rm B}$.
\item For the total range of O/H of the studied XMD galaxies of only 1.7
    (0.23 dex), the distribution of their global parameters vary widely.
   The range of $M$(\ion{H}{i}), as well as of blue luminosity, is two
   orders of magnitude (from $\sim$2$\times10^{7}$ to $\sim$3$\times10^{9}$
   M\sunn, and M$_{\rm B}$ from -12.5 to -17.7).
\item More than 2/3 of the twenty non-LSB XMD galaxies from this study show
    evidence of interaction/merger with other galaxies.
\item In the direction of AM~0624--261 ($l =$ 234.3, $b = -$16.8),
  an \ion{H}{i}-object is detected at a radial velocity $V_{\rm hel} = $
  +188~\kms\ ($V_{\rm lsr}$= +208~\kms), with an integrated \ion{H}{i} flux
  within the NRT beam of 0.77 Jy~\kms\ and $W_{50}$=37~\kms. A comparison
  with data on nearby high-velocity \ion{H}{i} clouds (HVC) shows
  that this object is probably a part of a new HVC (tentative name HVC
  234.3-16.8+208).
\end{enumerate}

\begin{acknowledgements}

We are grateful to the NRT Comit\'e de Programmes for time allocation for
this project. SAP is grateful to OdP and CNRS for support of his visits
to Observatoire de Paris where the main part of this work was performed.
SAP acknowledges the partial support from the RFBR grant No. 06-02-16617.
The Nan\c {c}ay Radio Observatory is the Unit\'e scientifique de Nan\c {c}ay
of the Observatoire de Paris, associated as Unit\'e de Service et de Recherche
No. B704 with the French Centre National de la Recherche Scientifique.
The Nan\c {c}ay Observatory also gratefully acknowledges the financial support
of the Conseil Regional de la R\'egion Centre in France.
We thank the anonymous referee for useful comments and suggestions that
allowed us to improve our presentation.
We acknowledge the use of the NED and LEDA databases to check the
information on the properties of the studied dwarf galaxies and their
environment, and the SDSS database for the use of photometry
for many of the studied galaxies and of spectral information on some objects
in their environment.

\end{acknowledgements}

\end{document}